\documentclass[]{aa}  
\usepackage{graphicx}
\usepackage{txfonts}
\usepackage{xcolor}
\usepackage{comment}
\usepackage[normalem]{ulem}
\usepackage{placeins}

\begin{document} 

\title{From simulations to observations. Methodology and data release 
   of mock TNG50 galaxies at $0.3 < z < 0.7$ for WEAVE-StePS}

\author{A.~Ikhsanova\inst{1}
        \and
        L.~Costantin\inst{2}
        \and
        A.~Pizzella\inst{1,3}
        \and
        E.~M.~Corsini\inst{1,3}
        \and
        L.~Morelli\inst{4, 5}
        \and
        F. R. ~Ditrani\inst{5,6}
        \and
        A.~Ferré-Mateu\inst{7, 8}
        \and
        L.~Gabarra\inst{9}
        \and
        M.~Gullieuszik\inst{3}
        \and
        C. P.~Haines\inst{4, 5}
        \and
        A.~Iovino\inst{5}
        \and
        M.~Longhetti\inst{5}
        \and
        A.~Mercurio\inst{10, 11, 12}
        \and
        R.~Ragusa\inst{11}
        \and
        P.~Sánchez-Blázquez\inst{13}
        \and
        C.~Tortora\inst{11}
        \and
        B.~Vulcani\inst{3}
        \and
        S.~Zhou\inst{5}
        \and
        E.~Gafton\inst{14}
        \and
        F.~Pistis\inst{15, 16, 17}
}

\institute{Dipartimento di Fisica e Astronomia ``G. Galilei'', 
        Universit\`a di Padova, vicolo dell'Osservatorio 3, 35122, Padova, Italy 
        \and
        Centro de Astrobiolog\'ia (CAB), CSIC-INTA, 
        Ctra. de Ajalvir km 4, 
        Torrej\'on de Ardoz, E-28850, Madrid, Spain
        \and
        INAF – Osservatorio Astronomico di Padova, vicolo dell’Osservatorio 5, 35122 Padova, Italy
        \and
        Instituto de Astronom\'ia y Ciencias Planetarias, Universidad de Atacama, Copiap\'o, Chile
        \and
        INAF-Osservatorio Astronomico di Brera, via Brera 28, I-20121 Milano, Italy
        \and
        Universit\`a degli studi di Milano-Bicocca, Piazza della Scienza, I-20125 Milano, Italy
        \and
        Instituto de Astrofísica de Canarias, IAC, Vía Láctea s/n, E-38205, La Laguna (S.C. Tenerife), Spain
        \and
        Departamento de Astrof\'isica, Universidad de La Laguna, E-38200, La Laguna, Tenerife, Spain
        \and
        Department of Physics, Oxford University, Keble Road, Oxford OX1 3RH, UK
        \and
        Università di Salerno, Dipartimento di Fisica “E.R. Caianiello”, Via Giovanni Paolo II 132, 84084 Fisciano (SA), Italy.
        \and
        INAF – Osservatorio Astronomico di Capodimonte, Salita Moiariello 16, 80131 Napoli, Italy
        \and
        INFN – Gruppo Collegato di Salerno – Sezione di Napoli, Dipartimento di Fisica “E.R. Caianiello”, Università di Salerno, Via Giovanni Paolo II, 132, 84084 Fisciano (SA), Italy.
        \and
        Departimento de F\'{\i}sica de la Tierra y Astrof\'{\i}sica , Fac. CC. F\'{\i}sicas \&  Instituto de F\'{\i}sica de Part\'{\i}culas y del Cosmos, IPARCOS, Universidad Complutense de Madrid, Plaza de las Ciencias 1, E-28040 Madrid, Spain 
        \and
        Isaac Newton Group of Telescopes, Apartado 321, 38700 Santa Cruz de la Palma, Tenerife, Spain
        \and
        Dipartimento di Fisica ``G. Occhialini'', Universit\`{a} degli Studi di Milano-Bicocca, Piazza della Scienza 3, I-20126 Milano, Italy
        \and
        National Centre for Nuclear Research, ul. Pasteura 7, 02-093 Warsaw, Poland
        \and
        INAF - Osservatorio di Astrofisica e Scienza dello Spazio di Bologna, Via Piero Gobetti 93/3, I-40129 Bologna, Italy
        \\
}

\date{Draft version: \today}

\abstract
{The new generation of optical spectrographs, 
i.e., WEAVE, 4MOST, DESI, and WST, 
with their huge multiplexing capabilities and
excellent spectral resolution, 
provide an unprecedented opportunity 
for statistically unveiling the details of the 
star formation histories of galaxies.
However, these observations
are not easily comparable with
predictions from cosmological simulations.}
{
Our goal is to build a reference
framework for comparing spectroscopic observations
with cosmological simulations and test currently 
available tools for deriving the stellar population 
properties of mock galaxies as well as 
their star formation histories.
}
{
In this work, we focus on the observational strategy of the Stellar Population at intermediate redshift Survey (StePS) carried out with the WEAVE instrument.
In particular, we create mock datasets of $\sim750$ galaxies at redshift $z=0.3$, $0.5$, and $0.7$
from the TNG50 cosmological simulation.
We perform radiative transfer calculations using \texttt{SKIRT} and analyze the spectra with the \texttt{pPXF} algorithm as if they are real observations.
}
{This work presents the methodology used to
generate the mock datasets and provide
an initial exploration of stellar population properties
(i.e., mass-weighted ages and metallicities) and
star formation histories of a test sample of three galaxies  
at $z=0.7$ and their descendants at $z=0.5$ and $0.3$.
We show that there is very good agreement between mock WEAVE-like spectra compared 
to the intrinsic values in TNG50 (average difference of $0.2\pm0.3$ Gyrs).
We also report that there is an overall agreement  
in retrieving the star formation history of galaxies,
especially if they form the bulk of their stars 
on short timescales and at early epochs.
While we find a tendency to overestimate the weight of old 
stellar populations in galaxies with complex star 
formation histories,
we properly recover the timescale on which galaxies build up 
$90\%$ of their mass
with almost no difference in the measured and intrinsic cumulative star formation histories over the last 4 Gyrs.}
{We release the datasets with this publication,
consisting of multi-wavelength imaging and spectroscopic data
of $\sim750$ galaxies at redshift $z=0.3$, 0.5, and 0.7.
This work provides a fundamental bench-test for the forthcoming WEAVE observations, providing the community with realistic mock spectra of galaxies that can be used to test currently available tools for deriving first-order stellar populations parameters (i.e., ages and metallicities) as well as more complex diagnostics such as mass and star formation histories.}


\titlerunning{TNG50 mock datasets at $0.3 < z < 0.7$}
\authorrunning{Ikhsanova et al.}

\maketitle

\section{Introduction}

Large extragalactic campaigns, like the Sloan Digital 
Sky Survey \citep[SDSS;][]{York.D:2000}, enriched our knowledge 
about the different galaxy populations in the nearby Universe
\citep[e.g.,][]{Strateva.I:2001, Kauffmann.G:2003, Blanton.M:2003, Baldry.I:2004, Gallazzi.A:2005, Gallazzi.A:2006, Thomas.D:2010, Goddard.D:2017}.
These observations revealed the existence of a bimodality 
in the properties of local galaxies
(i.e., morphology, color, spectral type, and star formation)
with red, quiescent, early-type galaxies separating themselves
from blue, star-forming, late-type galaxies. 
For example, galaxies in the \emph{blue cloud} 
of the color-magnitude diagram 
show a tight correlation between 
the integrated star formation rate ($SFR$) and stellar mass $M_{\star}$,
the so-called star formation main sequence 
\citep[SFMS; e.g.,][]{Brinchmann.J:2004, Daddi.E:2007, Whitaker.K:2012, Rinaldi.P:2024}, 
while galaxies in the \emph{red sequence} exhibit no $M_{\star}-SFR$ correlation.
These two populations are separated by 
transition galaxies, lying in the 
so-called \emph{green valley} \citep{Salim.S:2007}. 
This segregation of galaxies reflects the heterogeneity 
of physical processes affecting their evolution.
Indeed, the comparison of galaxy properties at different 
redshifts show that they could experience 
transformations over cosmic time, mostly due to 
the cessation of star formation activity and rejuvenation events
\citep{Lilly.S:2013, Vulcani.B:2015, 
Mancini.C:2019, ForsterSchreiber.N:2020, PerezGonzalez.P:2023, 
LeBail.A:2024, Carnall.A:2024}.

Our ability to reconstruct the star formation histories (SFHs) of galaxies from 
their stellar populations is key to tracing their individual evolution and constraining their transformation. In this context, a common approach 
consists of \emph{archaeologically} 
reconstructing the SFHs of local galaxies, analyzing key features in their spectra 
and the possible relation with other galaxy properties, 
e.g., stellar mass, kinematics, 
star formation activity, and environment.
But, despite the large availability of 
high-resolution spectra of nearby galaxies,
it is still extremely difficult to resolve their 
complex SFHs, due to the degeneracy between age 
and metallicity and the slow evolution of low-mass stars in evolved stellar populations (mass-weighted ages > 5~Gyr), 
typical of the majority of massive galaxies 
\citep[$M_{\star}>10^{10}$~M$_{\odot}$;][]{Gallazzi.A:2005, Serra.P:2007}.

The star formation histories at early times can be better constrained at intermediate redshifts, 
when stellar populations were younger and galaxies contained more massive stars.
To date, the Large Early Galaxy Astrophysics Census Survey \citep[LEGA-C;][]{vanderWel.A:2016} 
is the best campaign suited to trace SFHs in individual galaxies at intermediate 
redshift ($0.6 < z < 1.0$), targeting approximately 3000 spectra 
with median signal-to-noise ratios $S/N\sim 20~\AA^{-1}$ \citep{Straatman.C:2018}.
This effort will soon be complemented by the Stellar Population at intermediate redshift 
Survey \citep[StePS;][]{Iovino.A:2023}, one of the eight surveys that will be carried 
out with the WHT Enhanced Area Velocity Explorer \citep[WEAVE;][]{Jin.S:2024}, the new wide-field 
spectroscopic facility for the 4.2-m William Herschel Telescope (WHT) in the Canary Islands.
WEAVE-StePS aims to obtain relatively high-resolution 
($R \sim 5000$) spectra with typical signal-to-noise ratios $S/N\sim 10~\AA^{-1}$ for a sample of $\sim25,000$ galaxies with magnitude
$m_i$ $\leq$ 20.5~mag, the majority of which lie in the redshift range $0.3<z<0.7$ \citep[for details, see][]{Iovino.A:2023}.
Thus, the survey will provide
reliable measurements of the emission lines and absorption 
features in the stellar continuum across a broad wavelength coverage $3660-9590$~$\AA$ for a large sample of galaxies in different environments.
As an additional effort, high-quality spectra 
($S/N\sim30~\AA^{-1}$) for more than 3000 galaxies
will be observed with the 4-metre Multi-Object Spectrograph Telescope (4MOST)
facility mounted on the  Visible and Infrared Survey Telescope for Astronomy (VISTA)
as part of the 4MOST-StePS campaign
\citep{Iovino.A:2023Msngr}.

This work is driven by the need for a better
understanding and characterization of complex
physical processes involved in galaxy evolution,
which are key ingredients of state-of-the-art
cosmological simulations. Moreover, properly comparing
the properties of observed and simulated galaxies 
is becoming crucial for both validating the physics implemented in the models and in designing the best observational strategies in extragalactic campaigns.
In this context, a powerful tool for investigating the synergy between 
observations and theory is the so-called \emph{forward 
modeling} of data \citep[e.g.,][]{Snyder.G:2015, RodriguezGomez.V:2019, Costantin.L:2023, Baes.M:2024, Euclid:2025},
which consists of generating and analyzing mock 
datasets from hydrodynamic simulations.
In this work, we create mock datasets 
mimicking WEAVE-StePS observations, but
the same strategy could be also used to 
create mock observations (i.e., accounting for noise and systematics)
that mimic any observational setup and
provide crucial constraints on the expected 
performance of any given facility.
These datasets are valuable not only for an apple-to-apple
comparison between observations and simulations, 
but also offer the possibility for simulated-based inference
of different pathways of galaxy evolution, which are only 
accessible following galaxies evolution across cosmic time. 

In this work, we present the methodology for creating
realistic mock observations from the TNG50 simulation
tailored for WEAVE-StePS.
In particular, we present the mock noiseless datasets
and briefly describe how to account for observational
conditions typical of WEAVE-StePS,
which will be fully characterized in a companion paper.
We test currently available tools for deriving 
the stellar population properties 
of mock galaxies as well as their star formation histories.
We compare the measured age and metallicity 
of nine example galaxies with the intrinsic values
derived from TNG50. Moreover, we measure the
differences between their central and
integrated properties, looking for possible 
spatial variations.

The paper is organized as follows.
In Sect.~\ref{sec:section2}, we describe the 
main properties of the TNG50 cosmological 
simulation and the sample selection.
In Sect.~\ref{sec:section3}, we describe the methodology
to create the mock datasets.
In Sect.~\ref{sec:section4}, we present and discuss our
results, while in Sect.~\ref{sec:section5} we
summarize the main conclusions of this work.

Throughout this work, we assume a \citet{Planck2016} 
cosmology with
$H_0 = 67.74$~km~s$^{-1}$~Mpc$^{-1}$, $\Omega_{\rm m} = 0.3089$, 
and $\Omega_{\Lambda} = 0.6911$. 
We quote magnitudes in the AB system \citep{Oke.J:1983}. All errors are reported as the standard deviation unless stated otherwise.

\section{Data and sample \label{sec:section2}}

Motivated by the need for a proper comparison between 
cosmological simulations and upcoming 
observations with the WEAVE instrument, we build 
several mock datasets based on the TNG50-1 simulation and
tailored for WEAVE-StePS observations of 
massive galaxies ($M_{\star} > 10^{10.2}~$M$_{\odot}$)
at redshift $0.3 < z < 0.7$.

\begin{table*}[!h] 
\caption{Physical properties of selected sample of galaxies.} 
\label{tab:table1}      
\centering                          
\begin{tabular}{c c c c c c}       
\hline\hline                
$z$    & Number of galaxies  & $\log(M_{\rm total})$  & $\log(M_{\star})$     & $R_{\star}$  & $SFR$            \\
            &                     &  [M$_{\odot}$]  & [M$_{\odot}$] &[kpc]         & [M$_{\odot}$ yr$^{-1}$]   \\
(1) & (2) & (3) & (4) & (5) & (6) \\
\hline 
\vspace{0.2cm}
0.3         &   562     &  $11.91_{-0.45}^{+0.47}$  &   $10.65_{-0.27}^{+0.43}$    &  $4.66_{-2.50}^{+3.17}$    &   $3.82_{-3.43}^{+4.95}$   \\ 
\vspace{0.2cm}
0.5         &   155     &  $12.39_{-0.27}^{+0.40}$  &   $11.08_{-0.15}^{+0.41}$    &  $5.54_{-2.36}^{+3.41}$    &   $8.75_{-7.20}^{+11.23}$   \\
\vspace{0.2cm}
0.7         &   47      &  $12.8_{-0.21}^{+0.42}$  &   $11.47_{-0.21}^{+0.22}$    &  $5.83_{-2.01}^{+3.34}$    &   $17.06_{-8.04}^{+27.25}$   \\  
\hline        
\end{tabular}
\tablefoot{(1) Redshift. (2) Total number of galaxies in
the sample. (3) Median total mass of the subhalos. 
(4) Median stellar mass of the subhalos.
(5) Median 3D half-mass stellar radius. 
(6) Median star formation rate. Errors are quoted as 16th-84th percentile range.}
\end{table*}

TNG50 is a state-of-the-art cosmological 
magneto-hydrodynamical galaxy formation simulation
from the IllustrisTNG project
\citep{Weinberger.R:2017, Pillepich.A:2018}.
The simulation follows the evolution of $2 \times 2160^3$
total initial resolution elements within a uniform 
periodic-boundary cube of 51.7 comoving Mpc per side.
The spatial scale is set by a gravitational softening 
of dark matter and stars to be 0.575 comoving kpc until $z = 1$
and it is then fixed to its physical value of 
288~pc at $z = 1$, down to $z=0$. 
We refer to \citet{Pillepich.A:2019} and \citet{Nelson.D:2019} for all details about the
sub-grid physics implemented in the simulation.

We follow the WEAVE-StePS selection criteria as described in
\citet{Iovino.A:2023} and select TNG50 galaxies
in three redshifts snapshots (i.e., $z = [0.3, 0.5, 0.7]$),
having $\log(M_{\star}$/M$_{\odot}) > [10.28, 10.88, 11.23]$ and being brigher than $m_i = 20.5$~mag.
The sample consists of 764 galaxies, of which 562 are at $z=0.3$, 155 at $z=0.5$, and 47 at $z=0.7$. In our sample, we allow for duplicates, meaning the same galaxy could appear in different redshift snapshots.
In Table~\ref{tab:table1}, we summarize the distribution of their 
main physical properties.

As an example of our data products, we select three main progenitor branches from the TNG50 merger trees, ensuring that galaxies at redshift $z=0.7$ occupy different position in the $SFR-M_{\star}$ diagram, and follow their evolution across cosmic time. The physical properties (i.e., stellar mass, half-mass radius, and global $SFR$ of the three progenitors and their descendants) are detailed in Table~\ref{tab:table2}.

\begin{figure}[!h] 
\centering
\includegraphics[width=0.48\textwidth]{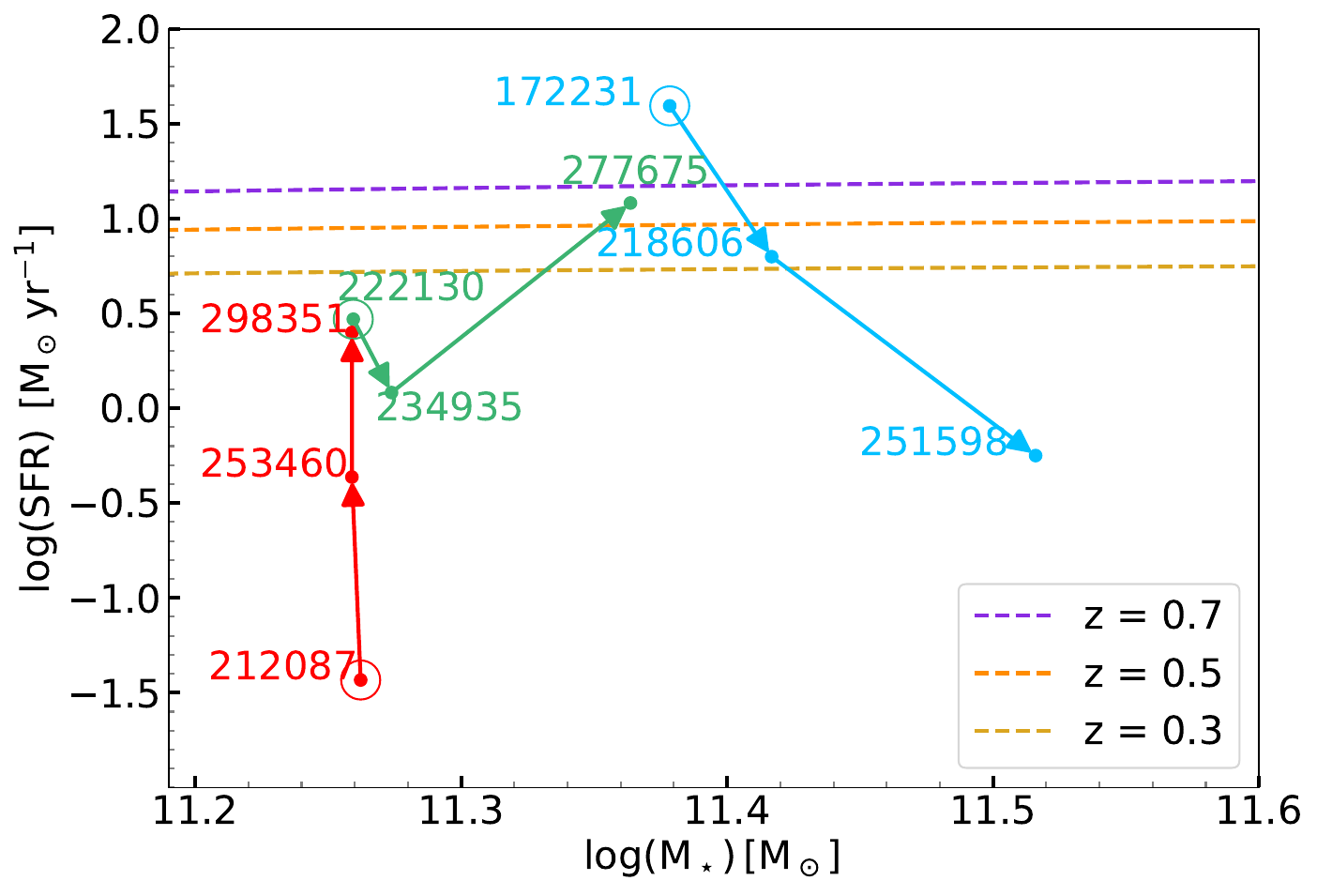}
\caption{
$SFR-M_{\star}$ time evolution of the 
three example galaxies (bigger symbols), from redshift $z=0.7$ to $0.3$.
The galaxy ID follows the nomenclature as in TNG50.
Galaxy ID~172231 (blue arrows) slowly quenches 
(ID~218606) until it reaches the red sequence (ID~251598).
Galaxy ID~222130 (green arrows) stays in the green valley 
(ID~234935) and then moves to the blue cloud sequence (ID~277675).
Galaxy ID~212087 (red arrows) slowly rejuvenates 
(ID~253460) and then moves to the green valley (ID~298351). 
Dashed lines mark the main sequence at z=0.3, 0.5, and 0.7 as defined in \citet{Koprowski.M:2024}.
\label{fig:figure1}}
\end{figure}

Figure~\ref{fig:figure1} shows the evolution of $SFR$ and stellar mass of
these three galaxies as a function of redshift.
We separate galaxies into the star-forming, 
green-valley, or quiescent categories following the criteria in \citet{Croom.S:2021} and \citet{Vaughan.S:2022}.
Galaxy ID~172231 is highly star forming at $z=0.7$ and 
it slowly quenches, evolving into ID~218606 at $z=0.5$
(star forming) and into ID~251598 at $z=0.3$
(green valley).
Galaxy ID~222130 (green valley) evolves into ID~234935 at $z=0.5$
(green valley) and into ID~277675 at $z=0.3$
(star forming).
Galaxy ID~212087 (quiescent) slowly rejuvenates, evolving into ID~253460 at $z=0.5$
(green valley) and into ID~298351 at $z=0.3$
(star forming). These three examples provide excellent test cases accounting for the variety of complex SFH and the many paths of galaxy evolution.
For our test sample, it is worth noting that the descendants 
share the same SFH of their progenitors,
since no major events happens in the evolution
of the three example galaxies from $z=0.7$ to $0.3$.

\begin{table}[!h] 
\caption{Physical properties of three examples of galaxies and their descendants.
}
\label{tab:table2}      
\centering                          
\begin{tabular}{c c c c c c}        
\hline\hline                 
$z$         & ID     &  Class   & $\log(M_{\star})$        & $R_{\star}$   & $SFR$                     \\
            &        &          & [M$_{\odot}$]  & [kpc]         & [$M_{\odot}$ yr$^{-1}$]   \\
(1)         & (2)    & (3)      & (4)               & (5)           & (6)            \\
\hline
0.7         & 172231 & SF       &     11.38         &  5.83         &  39.29         \\ 
0.5         & 218606 & SF       &     11.42         &  5.69         &  6.30          \\ 
0.3         & 251598 & GV       &     11.52         &  8.38         &  0.56          \\ 
\hline    
0.7         & 222130 & GV       &     11.26         &  4.84         &  2.95          \\ 
0.5         & 234935 & GV       &     11.27         &  5.49         &  1.21          \\ 
0.3         & 277675 & SF       &     11.36         &  7.51         &  12.08         \\ 
\hline   
0.7         & 212087 & Q        &     11.26         &  1.32         &  0.04          \\ 
0.5         & 253460 & GV       &     11.26         &  1.41         &  0.43          \\ 
0.3         & 298351 & SF       &     11.26         &  1.48         &  2.51          \\ 
\hline        
\end{tabular}
\tablefoot{(1) Redshift. (2) Galaxy ID. (3) Class: SF = star forming, 
GV = green valley, Q = quiescent  \cite[see][]{Croom.S:2021}.
(4) Stellar mass within half-mass radius. 
(5) Half-mass stellar radius. 
(6) Global star formation rate.}
\end{table}

\begin{figure*}[!h] 
\centering
\includegraphics[width=\textwidth]{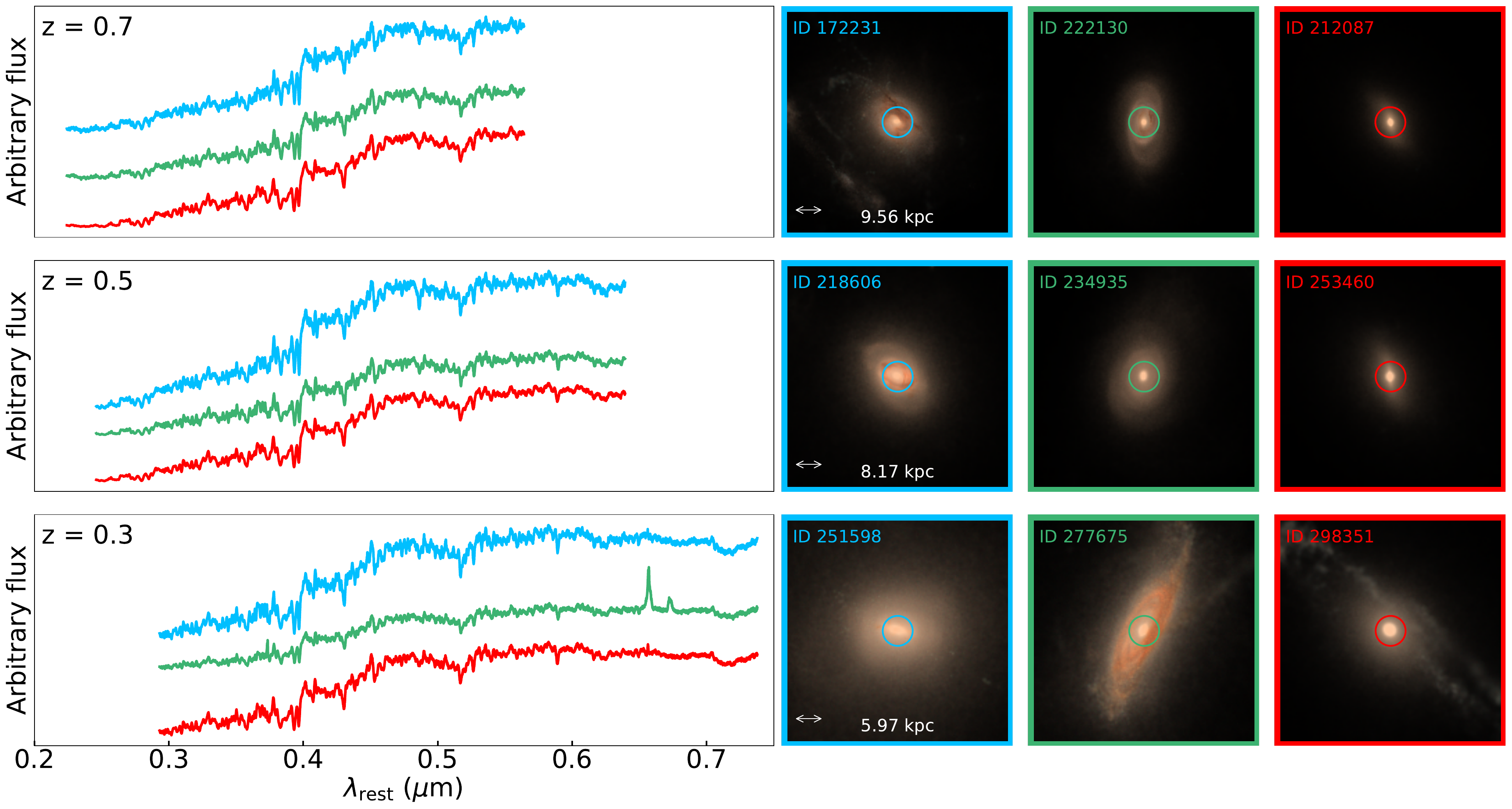}
\caption{Rest-frame spectra (first column) and
RGB images (second to fourth columns)
for three example galaxies at redshift $z=0.7$ (top row),
$z=0.5$ (middle row), and $z=0.3$ (bottom row),
respectively.
The first column demonstrates fiber spectra
extracted from the central 1.3~arcsec of each galaxy,
with arbitrary offset applied in the flux (linear units). 
In each panel of the first column: green spectra are extracted from
galaxy ID~222130, ID~234935, and ID~277675, respectively;
blue spectra from
galaxy ID~172231, ID~218606, and ID~251598, respectively;
red spectra from
galaxy ID~212087, ID~253460, and ID~298351, respectively.
The fiber size and corresponding
physical size are also shown.
\label{fig:figure2}}
\end{figure*}

Figure~\ref{fig:figure2} shows the 
RGB images of the target galaxies (right panels)
and their fiber noiseless spectra (left panels)
that we create following the procedure described in
Sect.~\ref{sec:section3}. 
For each set of galaxies, the different
rest-frame wavelength covered depends on their redshift.
In particular, spectra for galaxies at z = 0.7 reach the bluer part 
of the spectrum, covering absorption features such as 
\ion{FeII}2402, BL2538, \ion{FeII}2609, and \ion{MgII}.
On the other hand, in galaxies at $z=0.3$ we can target
the H$\alpha$ region, which provides the best proxy
of star-forming regions.
At all redshifts, the H$\beta$ and [\ion{OIII}]5007 emission lines 
are available.

It is worth mentioning that in this work 
we present the main methodology 
(Sect.~\ref{sec:section3}) and SFH analysis 
(Sects.~\ref{sec:section4.1} and ~\ref{sec:section4.2})
for three example galaxies at $z=0.7$ and their descendants
at $z=0.5$ and $0.3$,
discussing their mass-weighted age,
metallicity, and formation timescale.
While we have chosen to release the full suite of mock spectra with this current paper, our full analysis of the entire sample will be presented in a forthcoming companion paper (Ikhsanova et al.~\emph{in prep.}).

\section{Mock datasets \label{sec:section3}}

In this section we describe the strategy to create 
mock (realistic) spectra of galaxies starting 
from state-of-the-art cosmological simulations. 

In order to produce a noiseless mock spectrum, 
we retrieve the star and gas particles of
each selected galaxy in TNG50 and model the light distribution 
of the different stellar populations and star-forming regions, 
including the effects of dust on radiation.
For this task, we employ the \texttt{SKIRT}\footnote{\texttt{SKIRT} documentation: http://www.skirt.ugent.be.} 
Monte Carlo radiative transfer code 
\citep[v9.0;][]{Camps.P:2020},
which has been recently used to build 
mock datasets from different cosmological simulations
\citep{Costantin.L:2023, BarrientosAcevedo.D:2023, 
Baes.M:2024, Nanni.L:2024}. 

We follow the strategy described in \citet{Costantin.L:2023}, 
but focusing on a different redshift range.
For each galaxy and each snapshot (see Sect.~\ref{sec:section2}), 
we extract the corresponding sets of subhalo star particles 
and gas cells, while the stellar wind particles are ignored. 
As a note of caution, several assumptions are necessary 
to post-process numerical simulations (see e.g., Appendix~\ref{sec:appendixC}). 
In the following, we follow standard recipes \citep[i.e.,][]{Camps.P:2016, 
Camps.P:2018, Trayford.J:2017, RodriguezGomez.V:2019, 
Kapoor.A:2021}, but using a metallicity-dependent dust-to-metal ratio instead of a constant one \citep[see also][]{Popping.G:2022}.

\subsection{Primary source \label{sec:section3.1}}

For the primary source of emission, we assign 
to each star particle a spectral energy distribution 
(SED) that depends on its age and metallicity, which were extracted from the TNG50 database. 
To account for the galaxy kinematics, 
we include the velocity components of each star particle.
Each star particle older than 10~Myr 
is modeled with a simple stellar population (SSP)
from the BPASS library \citep[version 2.2.1;][]{Eldridge.J:2017, Stanway.E:2018} and a \citet{Chabrier.G:2003} 
initial mass function.
Each star particle younger than 10~Myr is
treated as an unresolved region of the interstellar medium 
(ISM) and is modeled with an SED from the 
MAPPINGS~III library \citep{Groves.B:2008},
which accounts both for the HII region and for the surrounding photodissociation region.

\subsection{Dust modeling \label{sec:section3.2}}

TNG50 does not include dust physics, hence we model 
the properties of dust in the ISM using the 
position, density, and metallicity of the Voronoi gas cells defined through the built-in 
Voronoi dust grid \citep{Camps.P:2013}.
Following \citet{Kapoor.A:2021}, the dust component is assumed 
to be traced by either star-forming ($SFR>0$) gas  cells 
or cold gas cells (i.e., with temperature $T<8000$~K).
The mass of the metals in the ISM is then converted
into dust mass with a dust-to-metal ratio 
using the relation described in \citet{Popping.G&Peroux.C:2022}.
To convert TNG50 metallicities to gas-phase metallicities, we assume that oxygen makes up $35\%$ of the metal 
mass and hydrogen $\sim 74\%$ of the baryonic mass.
The dust mass density of cold 
and star-forming gas cells is then set to be 
$\rho_{\rm dust} = f_{\rm dust}Z_{\rm gas}\rho_{\rm gas}$,
where $Z_{\rm gas}$ is the metallicity and $\rho_{\rm gas}$ is the density of the gas for cold and star-forming gas cells and zero for all the other gas cells. 

The dust composition is modeled with the dust mix of 
\citet{Zubko.V:2004}, which includes non-composite 
silicates, graphite, and neutral and ionized 
polycyclic aromatic hydrocarbon dust grains.
We allow dust grains to be stochastically heated 
and decoupled from local thermal equilibrium, 
also including dust self-absorption and re-emission.

\subsection{Mock data products \label{sec:section3.3}}

With the recipe described above, we create 
a set of noiseless mock data products.
The spectroscopic mock data set includes
both fiber spectra centered on the galaxy center
and integral field datacubes encompassing 
a field-of-view with diameter equal to
one half-mass radius of the dark matter subhalo.
In detail, each datacube has a spatial resolution of
0.4~arcsec~px$^{-1}$ and covers the observed spectral range
of WEAVE $ 3660-9590$~$\AA$ with a spectral sampling
of 1~\AA~px$^{-1}$. 
Then, each fiber spectrum is extracted from the integral-field datacube by considering only the spaxels within a circular aperture of 1.3~arcsec in diameter, mimicking WEAVE-StePS observations \citep[][]{Iovino.A:2023}.
In Figure~\ref{fig:figure2}, we present three showcase galaxies drawn from our sample at $z=0.7$ and their descendants at $z=0.5$ and $0.3$ (nine galaxies in total). These galaxies occupy different places in the $SFR-M_{\star}$ diagram (see also Fig.\ref{fig:figure1} and Sect.~\ref{sec:section2}).
Note that we follow the ID nomenclature as in TNG50, 
where galaxies can change ID in different snapshots.

In addition to the spectroscopic dataset, we defined a mock imaging dataset obtained by convolving \texttt{SKIRT} flux densities with the response curve for multiple filters. We create SDSS ($ugriz$) images with
spatial resolution of 0.396~arcsec~px$^{-1}$, 
ACS/WFC3 (F606W, F814W, F105W, F125W, F160W) images with
spatial resolution of 0.03 and 0.06 arcsec~px$^{-1}$, 
and NIRCam (F090W, F115W, F150W, F200W, 
F277W, F356W, F444W) images with
native spatial resolution of 0.031 and 0.063 arcsec~px$^{-1}$
for the short and long channel, respectively.
In Appendix~\ref{sec:appendixA}, we show the imaging dataset for galaxy ID~172231. 
As an additional product, 
we built composite RGB images using 
HST/WFC3 F435W, F606W, and F814W as red, green, and blue layers, respectively. 
An example of these composite images can be seen in Figure~\ref{fig:figure2}.
It is worth noting that this work focuses 
on the analysis of spectroscopic datasets, while the 
complementary imaging dataset will provide invaluable 
legacy value to further characterize the properties of
these galaxies and their morphological transformation.

\subsection{StePS-like data products \label{sec:section3.4}}

Since our goal is to reproduce realistic mock observations
of TNG50 galaxies, we mimic the observational 
strategy of WEAVE-StePS by perturbing 
the mock spectra to account for seeing, noise, 
and spectral resolution effects.
Firstly, we convolve each spectral slice of the mock datacube with a Gaussian function of $FWHM = 0.69$~arcsec, corresponding to the typical seeing for the WHT \citep{Wilson.R:1999}.
Then, we extract a spectrum within a circular aperture
centered on the galaxy with a diameter of 1.3~arcsec.
Finally, we follow the recipe described 
in \citet{Costantin.L:2019} and \citet{Ditrani.F:2023} 
and perturb each fiber spectrum accounting for 
the combined response curve 
of the WHT and that of the WEAVE spectrograph, 
Poisson contribution due to source and sky background, 
and readout noise of the WEAVE CCDs.
In particular, the sky noise is computed assuming typical conditions of dark nights in La Palma \citep[surface brightness $V \sim 22$ $\rm mag$ $\rm arcsec^{-2}$;][]{Benn.C:1998} , while the read-out noise was assumed to be $\sim2.5 \rm e^{-} \rm px^{-1}$ \citep{Dalton.G:2016}. Since StePS observing strategy consists of short exposures ($\sim$ 20 min), we neglect the contribution of thermal noise due to dark current ($< 0.1 \rm e^{-} h^{-1}$). Each noise contribution is added in quadrature to estimate the total noise, which is used as the $\sigma$ of a Gaussian distribution to perturb each normalized spectrum.
While in this work we focus on mass-weighted 
ages and metallicities derived from noiseless spectra
(Sect.~\ref{sec:section4.1}),
in a companion paper we will present the effect of 
$S/N$ in retrieving the main stellar population
properties of the entire sample.

\section{Results and discussion \label{sec:section4}}

In this section, we describe the methodology
to create and analyze mock observations of intermediate-redshift galaxies
tailored for WEAVE-StePS.
Firstly, we quantify possible
biases in observing galaxies with limited spatial information, looking at their central and integrated properties.
Then, we test currently available tools
for deriving the stellar population properties of mock galaxies and compare the measured
age and metallicity of nine example galaxies 
with the intrinsic values derived from TNG50.
Finally, we compare the timescale of star formation
from the mix of different stellar populations
both in simulations and observations.

\subsection{Mass-weighted age and metallicity \label{sec:section4.1}}

We retrieve the mass-weighted age and
metallicity of our example galaxies, comparing
the observed values with the intrinsic ones
obtained from the simulation. Furthermore,
we compare how these quantities vary if we analyze
the spectrum of the entire galaxy or just
the central 1.3~arcsec.

\subsubsection{TNG50 \label{sec:section4.1.1}}

We compare the age and metallicity of the entire 
subhalo in TNG50 with the values
obtained for the central aperture.
The latter is done by selecting all the star particles 
within a circular aperture of 1.3~arcsec.
We then weight the age and metallicity distribution of the star particles according to their mass.

\begin{table*}[!h] 
\caption{Mass-weighted ages and  metallicities of the example galaxies.
\label{tab:table3}} 
\centering
\begin{tabular}{c c c c c c c c c c c} 
\hline\hline 
$z$    & ID     & Class    & age$_{\rm sim}^{\rm total}$  & age$_{\rm sim}^{\rm fiber}$   & age$_{\rm obs}^{\rm total}$   & age$_{\rm obs}^{\rm fiber}$   & [Z/H]$_{\rm sim}^{\rm total}$  & [Z/H]$_{\rm sim}^{\rm fiber}$   & [Z/H]$_{\rm obs}^{\rm total}$   & [Z/H]$_{\rm obs}^{\rm fiber}$   \\
            &        &          & [Gyr]               & [Gyr]                           & [Gyr]            & [Gyr] & [dex] & [dex] & [dex] & [dex]       \\
(1) & (2) & (3)  & (4) & (5)  & (6)  & (7)  & (8)  & (9)  & (10)   & (11)   \\

\hline
0.7 & 172231 & SF & 3.29 & 3.39 & 3.8 $\pm$ 0.2 & 4.0 $\pm$ 0.2 & 0.08 & 0.26 & $-0.58$ $\pm$ 0.04 & $-0.56$ $\pm$ 0.04 \\
0.5 & 218606 & SF & 4.23 & 4.28 & 4.8 $\pm$ 0.3 & 4.6 $\pm$ 0.3 & 0.09 & 0.29 & $-0.13$ $\pm$ 0.02 & $-0.02$ $\pm$ 0.02 \\
0.3 & 251598 & GV & 5.87 & 6.31 & 6.3 $\pm$ 0.2 & 6.0 $\pm$ 0.3 & 0.09 & 0.31 & $-0.17$ $\pm$ 0.03 & 0.06 $\pm$ 0.03 \\
\hline
0.7 & 222130 & GV & 4.12 & 4.50 & 4.9 $\pm$ 0.2 & 4.7 $\pm$ 0.2 & 0.12 & 0.29 & $-0.06$ $\pm$ 0.02 & $-0.08$ $\pm$ 0.02 \\
0.5 & 234935 & GV & 5.30 & 5.94 & 5.7 $\pm$ 0.2 & 6.0 $\pm$ 0.3 & 0.10 & 0.31 & $-0.08$ $\pm$ 0.02 & 0.09 $\pm$ 0.02 \\
0.3 & 277675 & SF & 6.31 & 7.38 & 8.4 $\pm$ 0.4 & 8.0 $\pm$ 0.4 & 0.08 & 0.32 & $-0.47$ $\pm$ 0.05 & 0.02 $\pm$ 0.03 \\
\hline
0.7 & 212087 & Q & 5.38 & 5.53 & 5.5 $\pm$ 0.1 & 5.6 $\pm$ 0.1 & 0.13 & 0.28 & $-0.08$ $\pm$ 0.02 & 0.04 $\pm$ 0.01 \\
0.5 & 253460 & GV & 6.67 & 6.84 & 6.9 $\pm$ 0.2 & 7.0 $\pm$ 0.2 & 0.13 & 0.29 & $-0.08$ $\pm$ 0.01 & 0.04 $\pm$ 0.01 \\
0.3 & 298351 & SF & 8.28 & 8.59 & 8.5 $\pm$ 0.5 & 8.5 $\pm$ 0.2 & 0.12 & 0.31 & 0.03 $\pm$ 0.04 & 0.08 $\pm$ 0.01 \\

\hline                                  
\end{tabular}
\tablefoot{(1) Redshift. (2) Galaxy ID. (3) Class: SF = star forming, 
GV = green valley, Q = quiescent. 
(4) Mass-weighted age of the entire galaxy derived from TNG50.
(5) Mass-weighted age of the central 1.3~arcsec derived from TNG50.
(6) Mass-weighted age of the entire galaxy derived with \texttt{pPXF} from noiseless spectra.
(7) Mass-weighted age of the central 1.3~arcsec derived with \texttt{pPXF} from noiseless spectra.
(8) Mass-weighted metallicity of the entire galaxy derived from TNG50.
(9) Mass-weighted metallicity of the central 1.3~arcsec derived from TNG50.
(10) Mass-weighted metallicity of the entire galaxy derived with \texttt{pPXF} from noiseless spectra.
(11) Mass-weighted metallicity of the central 1.3~arcsec derived with \texttt{pPXF} from noiseless spectra.}
\end{table*}

In Table~\ref{tab:table3} (columns 4-5), it can be seen
that the average difference in the mass-weighted age
between the central region and
the entire galaxy is usually small (average of $\sim 0.4$~Gyr), 
although the core is systematically older
than the entire galaxy. This is consistent with
the majority of star formation taking place in
the disk while the bulge/nuclear region 
consists of a more mature stellar population,
in agreement with an inside-out growth 
\citep[e.g.,][]{Nelson.E:2021, Costantin.L:2021, Costantin.L:2022}.
A similar trend is observed in the metallicity 
difference (Table~\ref{tab:table3}, columns 8-9), 
with the central region being systematically
more metal-rich than the entire galaxy 
(average of $\sim 0.2$~dex).

It is worth noticing that, using a fixed aperture,
at $z=0.7$ we are probing larger 
physical scales ($9.56$ kpc) compared to $z=0.3$  ($5.97$ kpc).
Thus, the contamination from the mixed bulge and disk
populations are more significant at the highest redshifts.
The average age difference is $0.21 \pm 0.12$~Gyr at $z=0.7$ 
and $0.61 \pm 0.33$~Gyr at $z=0.3$.

\subsubsection{Mock observations \label{sec:section4.1.2}}

As for real galaxies, we analyze each mock spectrum 
with the Penalized Pixel-Fitting algorithm
\citep[\texttt{pPXF};][]{Cappellari.M:2017, Cappellari.M:2023} 
and derive the non-parametric SFH of each galaxy,
modeling both the stellar and gas components.
We fit the spectra with the BPASS library and Chabrier initial mass function \citep{Chabrier.G:2003},
consistently with our strategy outlined in Sect.~\ref{sec:section3.1}.
The wavelength grid has a resolution of 1~$\AA$ over the full range.
We consider models with stellar ages 
from 50~Myr 
to the age of the Universe at each considered redshift in 24, 23, and 22 bins at $z=0.3$, $0.5$, 
and $0.7$, respectively
and metallicities with the same setup as adopted in \citet{Cappellari.M:2023},
in 10 bins with [Z/H] = [$-1.3$, $-1$, $-0.8$, $-0.7$, 
$-0.5$, $-0.4$, $-0.3$, 0, 0.2, 0.3] dex.
It is worth noticing that the TNG50 star particles
may have metallicities above the upper limit of the templates,
as already discussed in \citet{Sarmiento.R:2023}.

We firstly run \texttt{pPXF} to measure the stellar kinematics 
(i.e., line of sight velocity and velocity dispersion) of each galaxy. Then, we run \texttt{pPXF} again, fixing the kinematics 
to the values obtained in the first iteration
to avoid degeneracies with the stellar 
population parameters 
\citep{SanchezBlazquez.P:2011, MartinNavarro.I:2024}.
If noisy spectra are analyzed, 
the noise should be scaled 
according to the residuals of 
the first fit to achieve a reduced $\chi^2_{\nu} = 1$. 
This scaling is required to explore various 
regularized solutions for the best-fitting weight distribution 
\citep[see][for more details]{Cappellari.M:2017, Cappellari.M:2023}, 
aiming to identify the optimal one, 
which was $\texttt{reg}\sim10-750$.
To better account for young stellar populations
and episodes of intense star formation,
we derive weights representing light fractions 
(e.g., we normalize the templates to 
the V-band prior to fitting)
and convert them to mass afterwards.
We do not apply any additive or multiplicative polynomial.

Following \citet{Kacharov.N:2018}, we perform a bootstrap analysis to estimate
the uncertainties in the distribution of the mass weights.
We sample each spectrum 100 times using the residuals 
of the best-fitting regularized solution.
Then, each spectrum is fitted with the same setup 
but without regularisation
to avoid self-similarity due to smooth solutions.

Following the procedure described above,
we fit both the integrated and StePS-like fiber spectra
of each galaxy. 
From the second \texttt{pPXF} run, 
we obtain the mean stellar age,
metallicity, and SFH of each galaxy
(Table~\ref{tab:table3}).
The results of the spectral fitting
for the three progenitor galaxies
(i.e., ID~172231, ID~212087, and ID~222130)
are shown in Fig.~\ref{fig:figureB}.

\begin{figure*}[!h]     
\centering
\includegraphics[width=6cm]{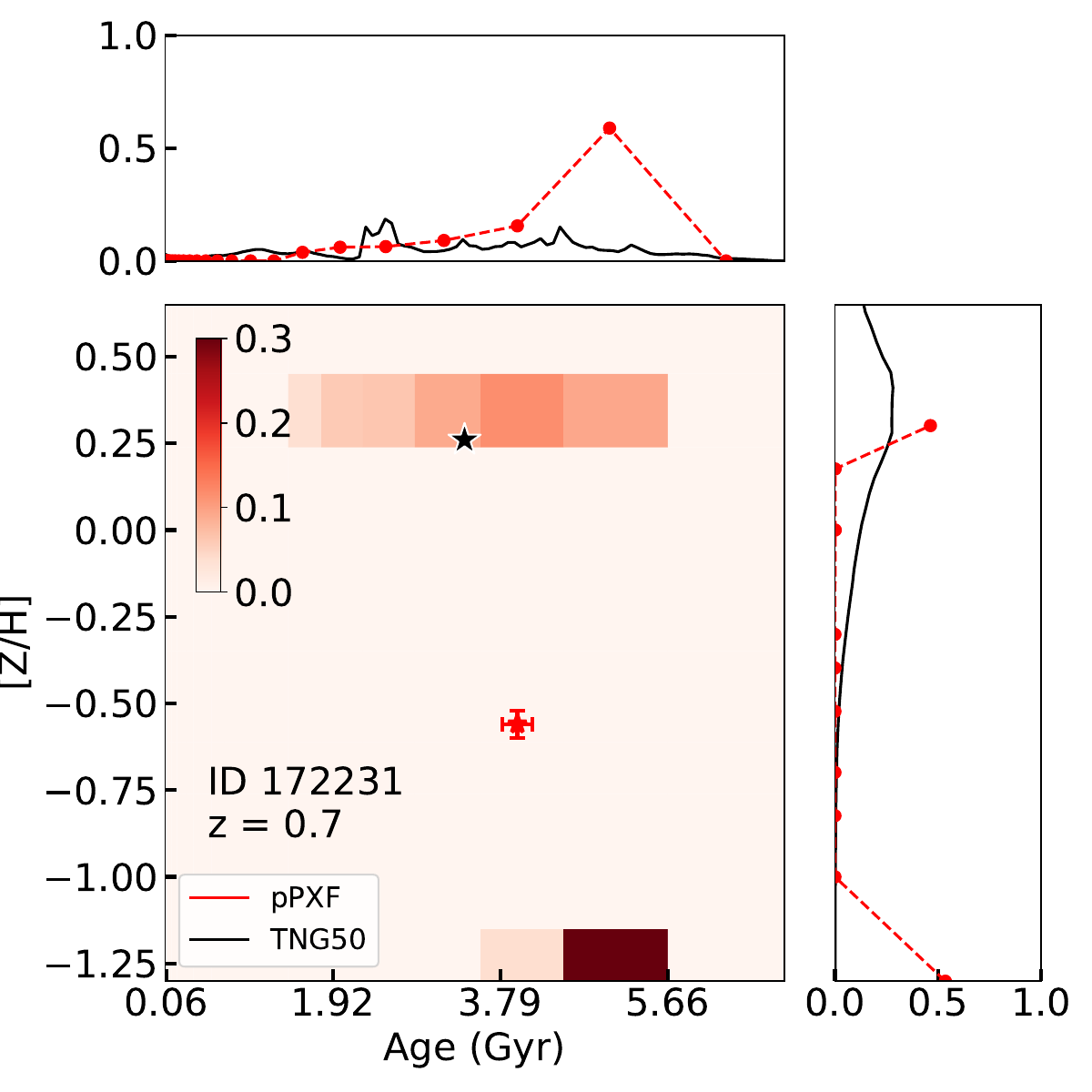}
\includegraphics[width=6cm]{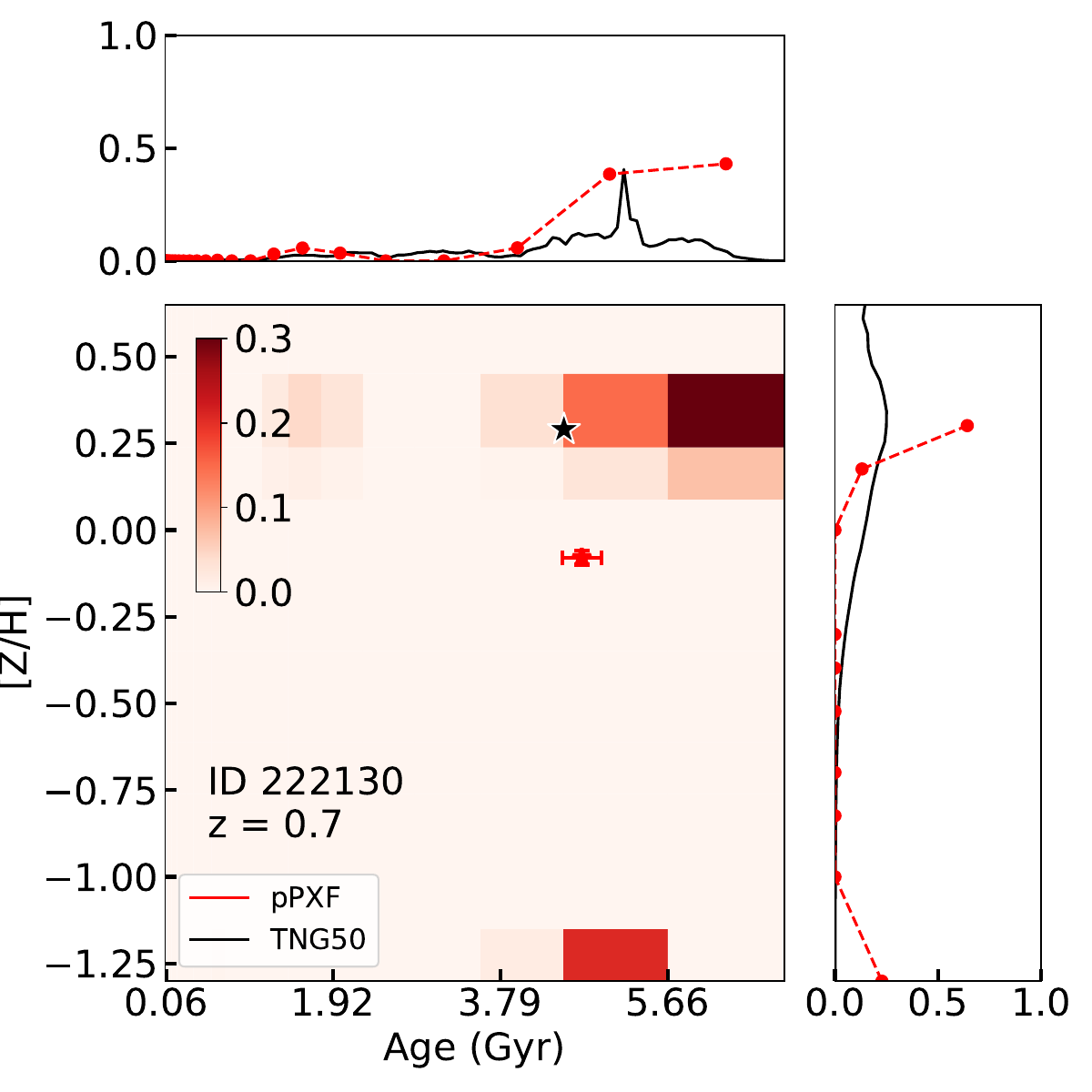}
\includegraphics[width=6cm]{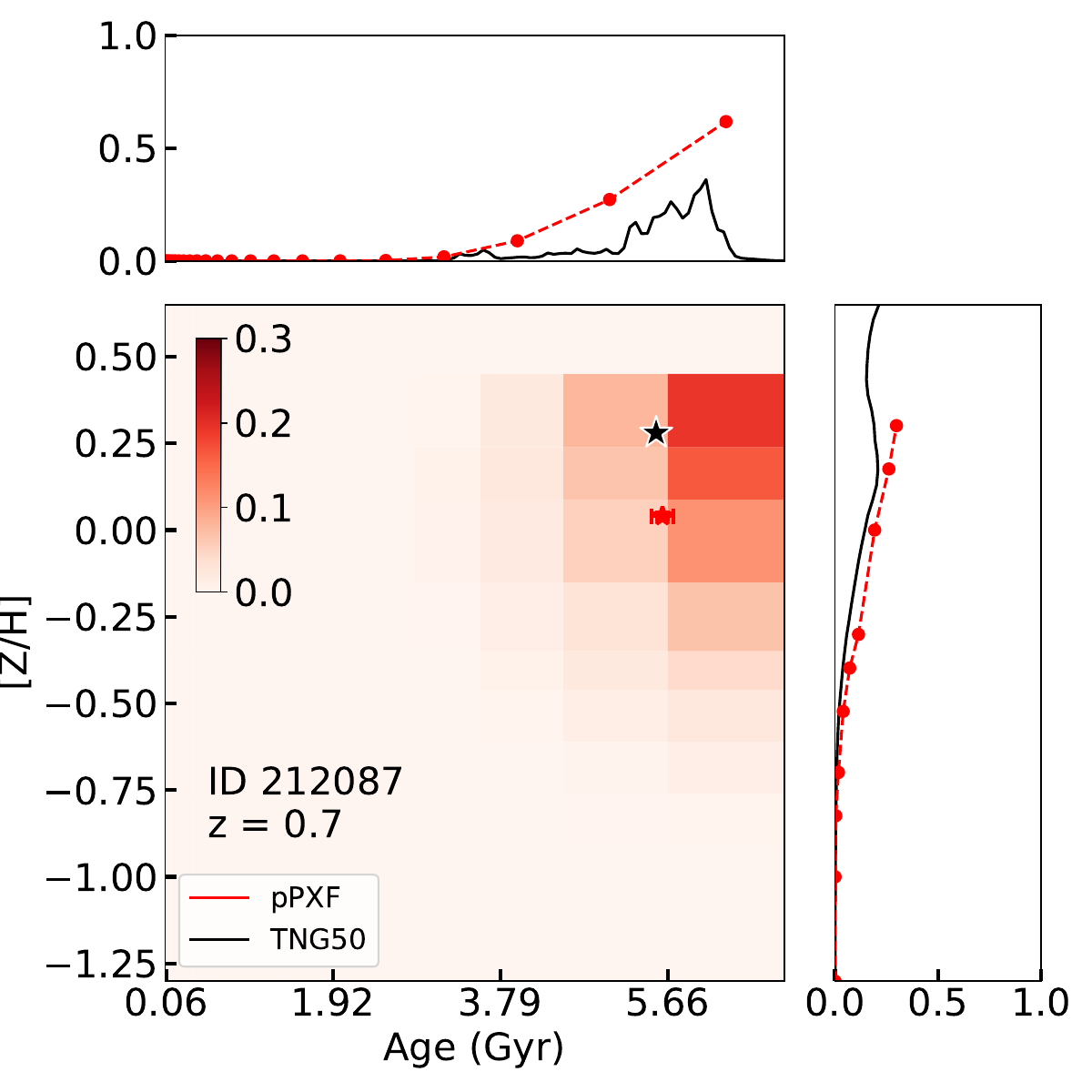}

\includegraphics[width=6cm]{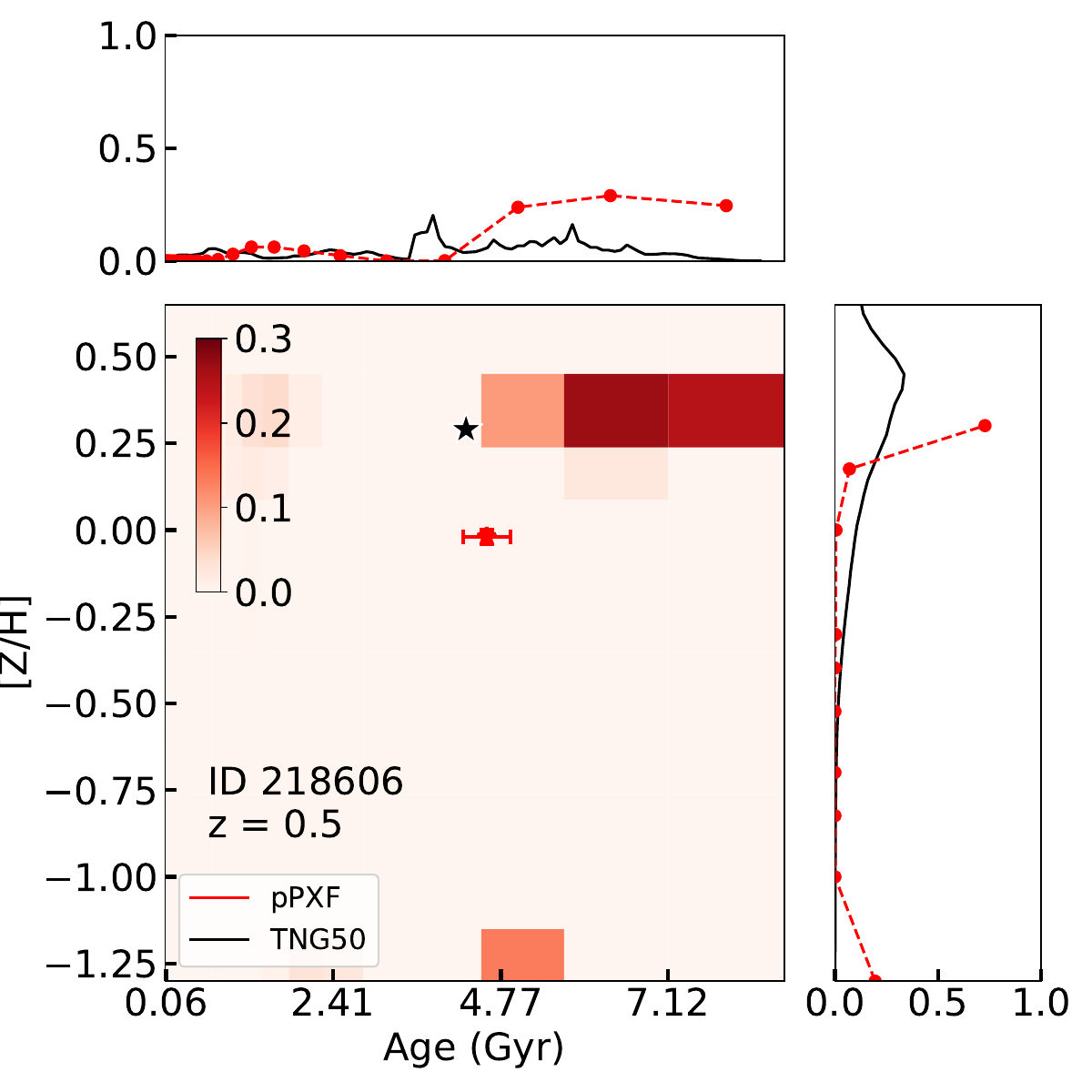}
\includegraphics[width=6cm]{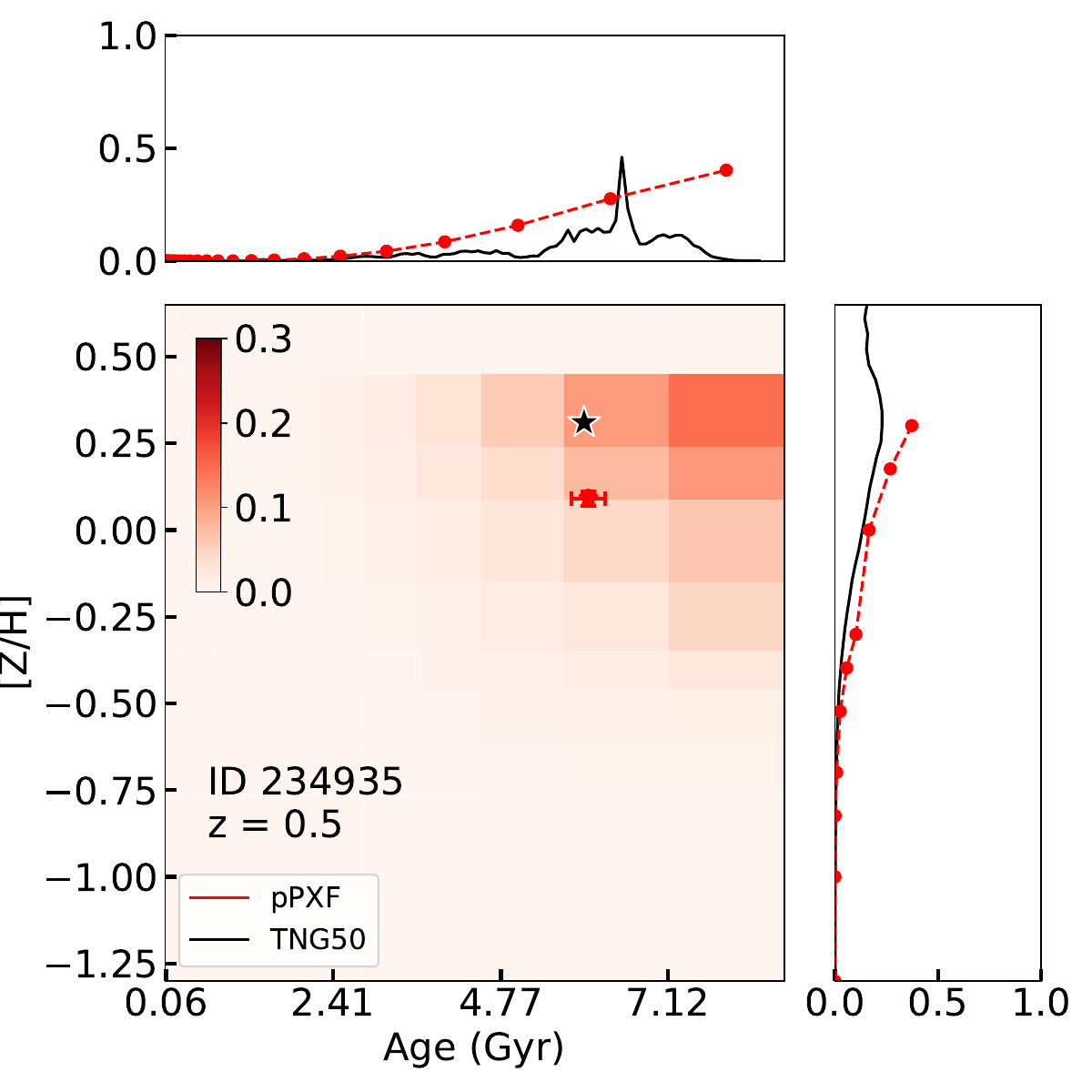}
\includegraphics[width=6cm]{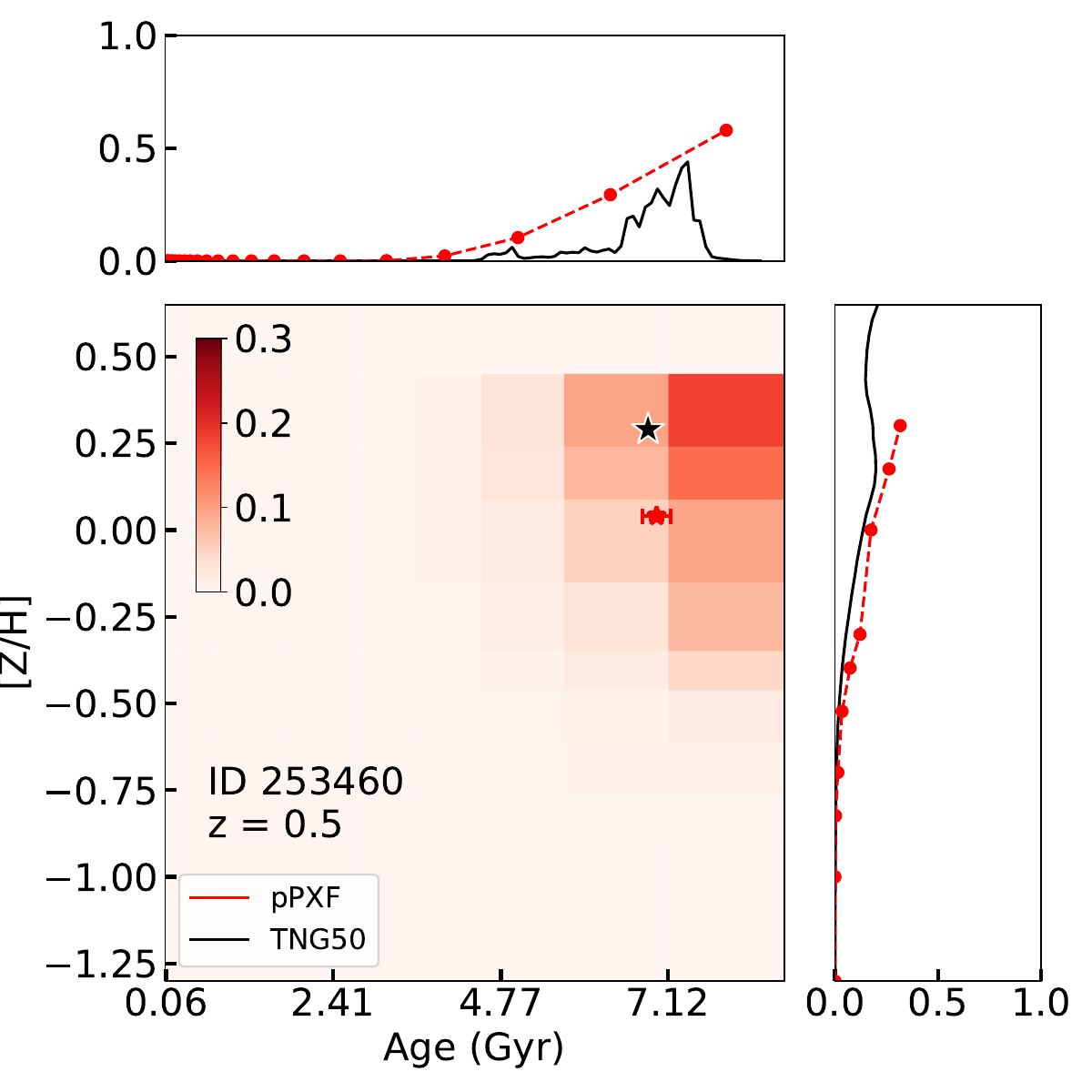}

\includegraphics[width=6cm]{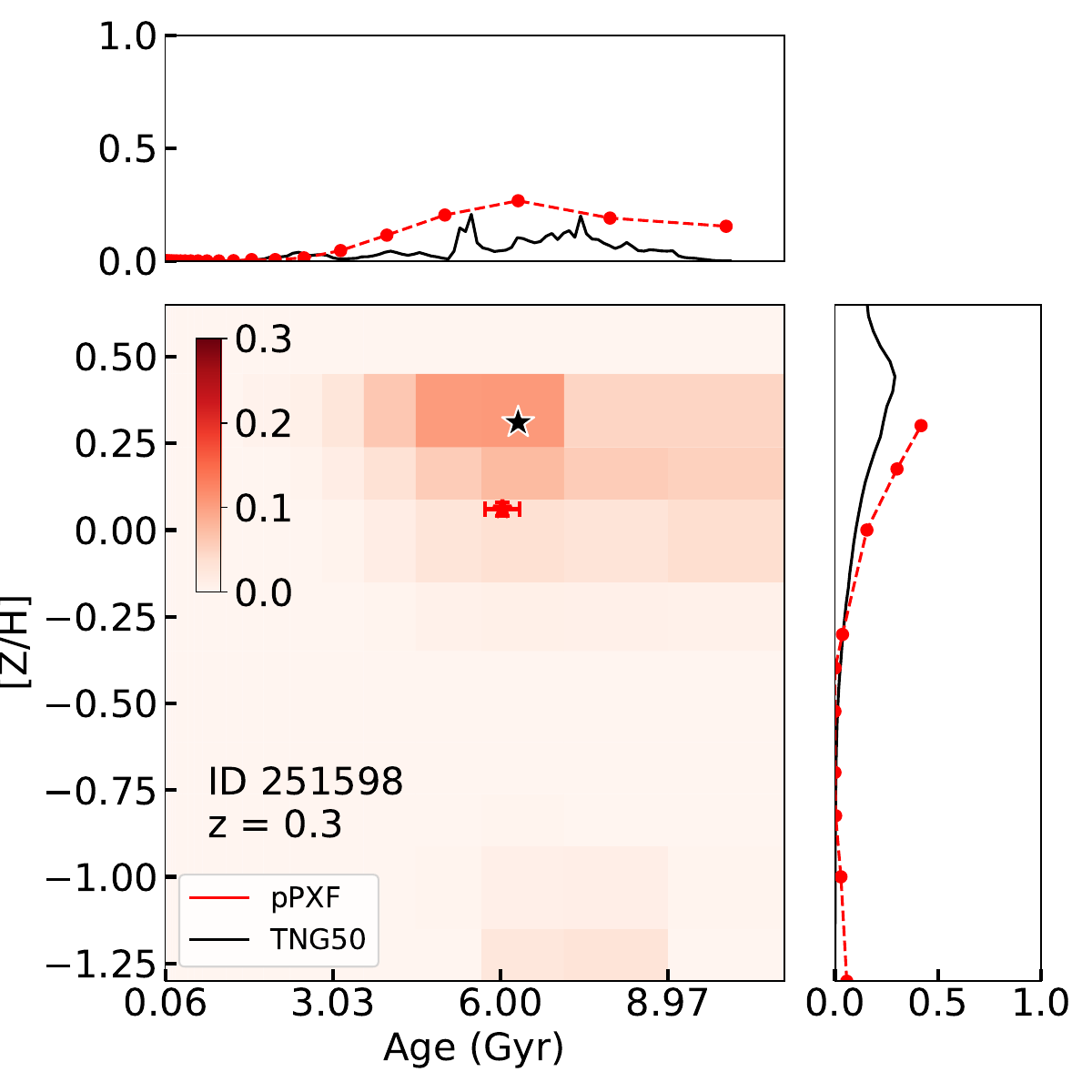}
\includegraphics[width=6cm]{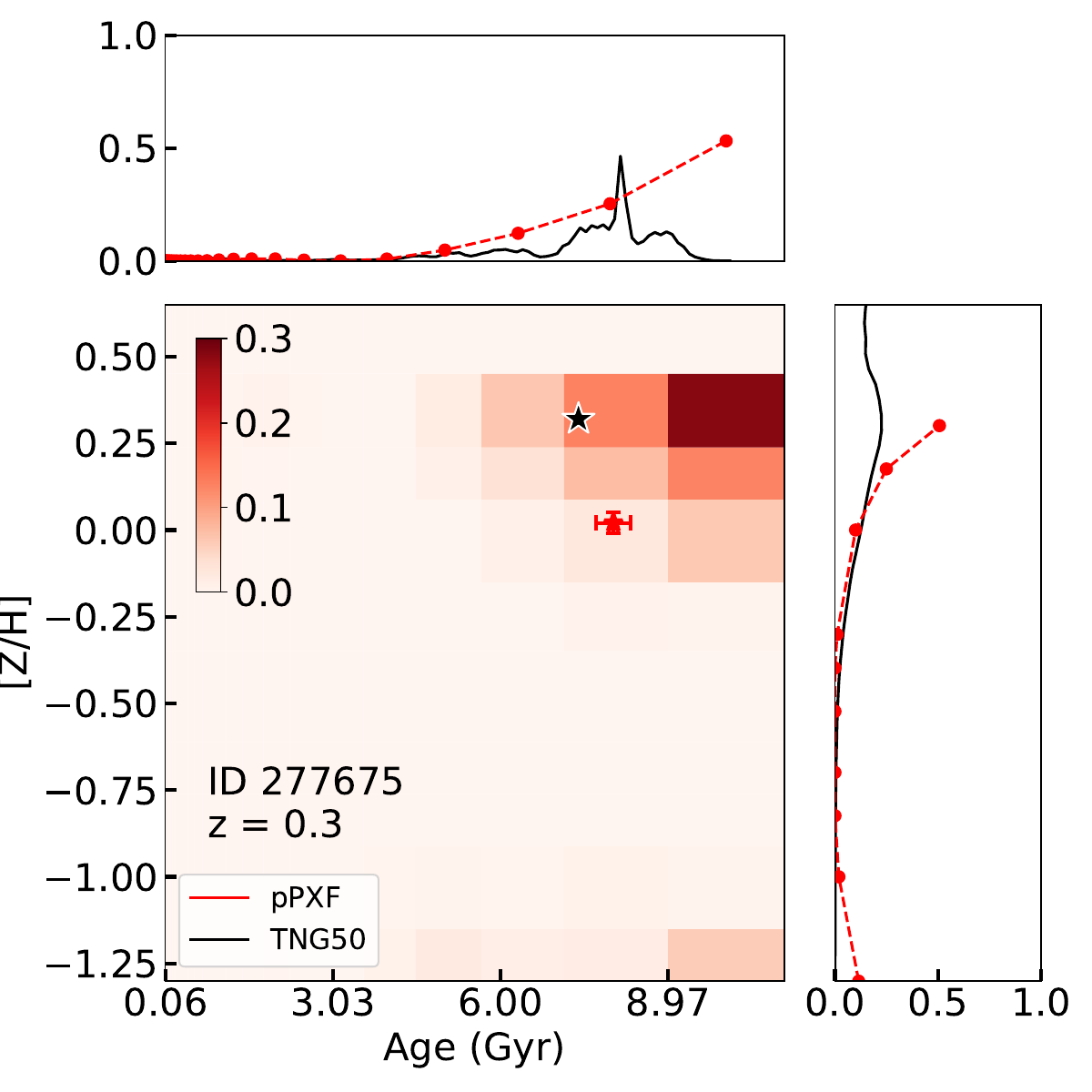}
\includegraphics[width=6cm]{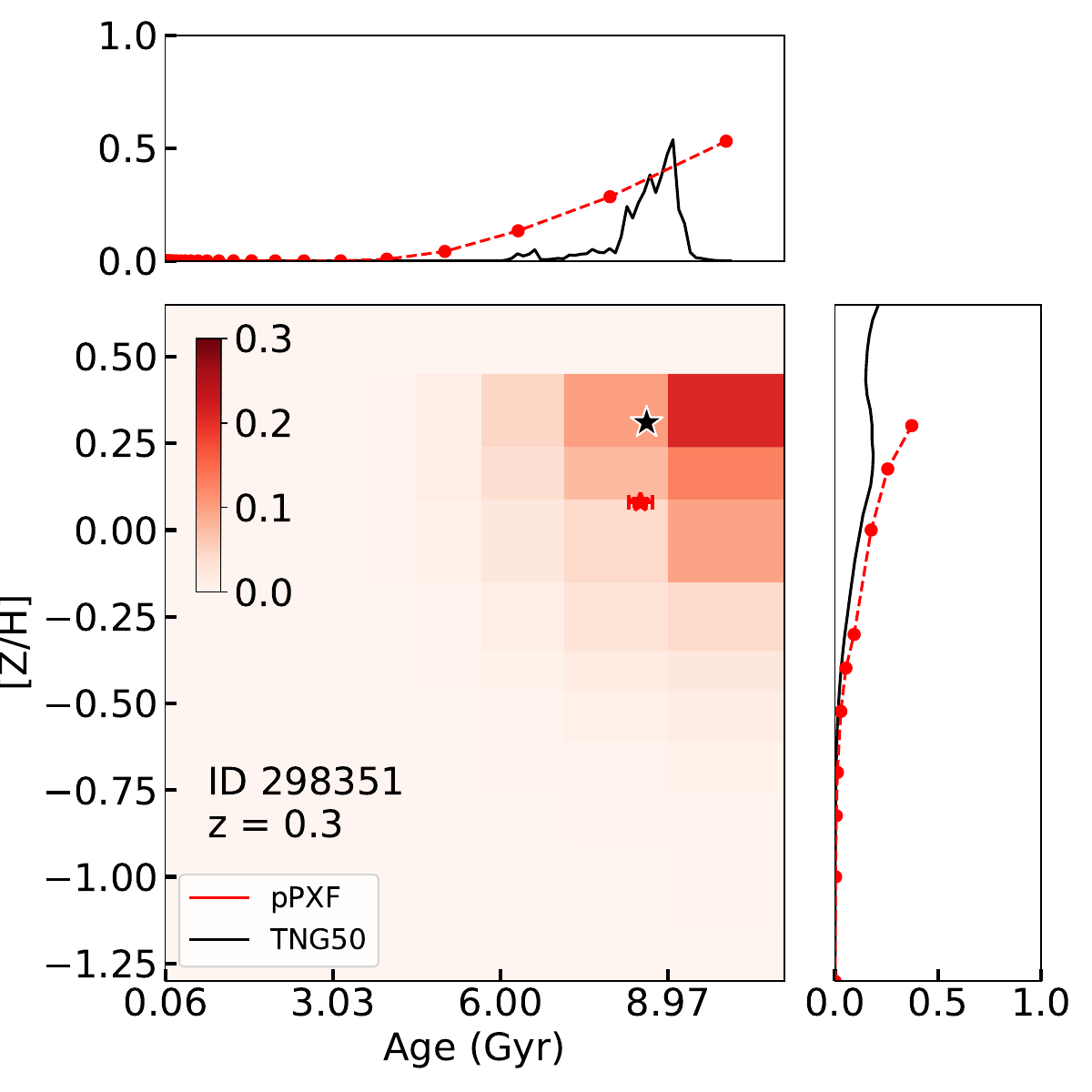}

\caption{Distribution of the \texttt{pPXF} weights (red colors), 
indicating the mass fraction of each stellar population 
of given age and metallicity, derived from
fiber noiseless spectra. 
The mass-weighted age and metallicity measured with
\texttt{pPXF} is shown as a red star, while
mass-weighted age and metallicity derived from the
simulation is marked as a black star.
Black lines represent the marginalized distribution
of ages and metallicities derived from the simulation,
while red curves show those inferred from \texttt{pPXF}.
The progenitor galaxies at $z=0.7$ are shown in the top row,
while their descendants at $z=0.5$ and $z=0.3$ are shown
in the middle and bottom rows, respectively.
\label{fig:figure3}}
\end{figure*}

In Figure~\ref{fig:figure3}, we compare 
the two-dimensional distribution of \texttt{pPXF} weights derived for the fiber noiseless spectra
in the age-metallicity diagram with the 
distribution of ages and metallicities 
derived from the simulation.
We find a good agreement in the mass-weighted ages
of the example galaxies.
For the fiber spectra, the values of age inferred from 
\texttt{pPXF} agree with the intrinsic ones, 
with an average difference of 
$0.2\pm0.3$~Gyr.
We find a similar trend for the integrated spectra, 
with measured ages compatible with intrinsic ones 
(average difference of $0.6\pm0.6$~Gyr.
Regarding metallicities, we find that measured values 
are systematically lower than the intrinsic ones,
with the caveat that
we are unable to retrieve 
metallicities as high as ${\rm [Z/H]}\sim0.3$ dex.
The average difference is
$0.3 \pm 0.2$~dex
between the fiber and integrated spectra. 

Overall, our analysis suggests that 
there is a good agreement in the inference of
mass-weighted ages, even if we
tend to predict lower metallicities for the galaxies,
as can also be seen in the marginalized
age and metallicity distribution (Fig.~\ref{fig:figure3}).
In a companion paper, we aim at quantifying 
statistically these specific trends for the selected
population described in Sect.~\ref{sec:section2},
also considering the effect of S/N in observations.

\subsection{Star formation history \label{sec:section4.2}}

\begin{figure*}[!h]     
\centering
\includegraphics[width=0.365\textwidth]{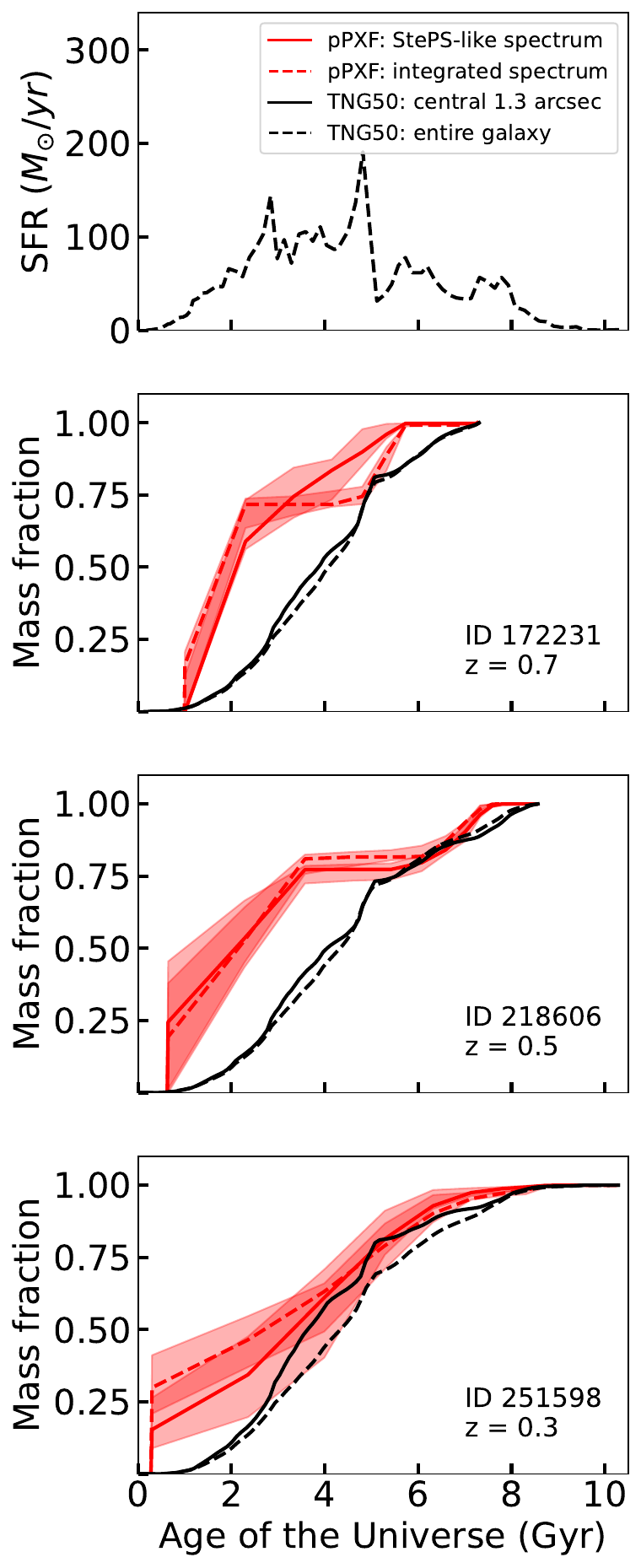}
\includegraphics[width=0.30\textwidth]{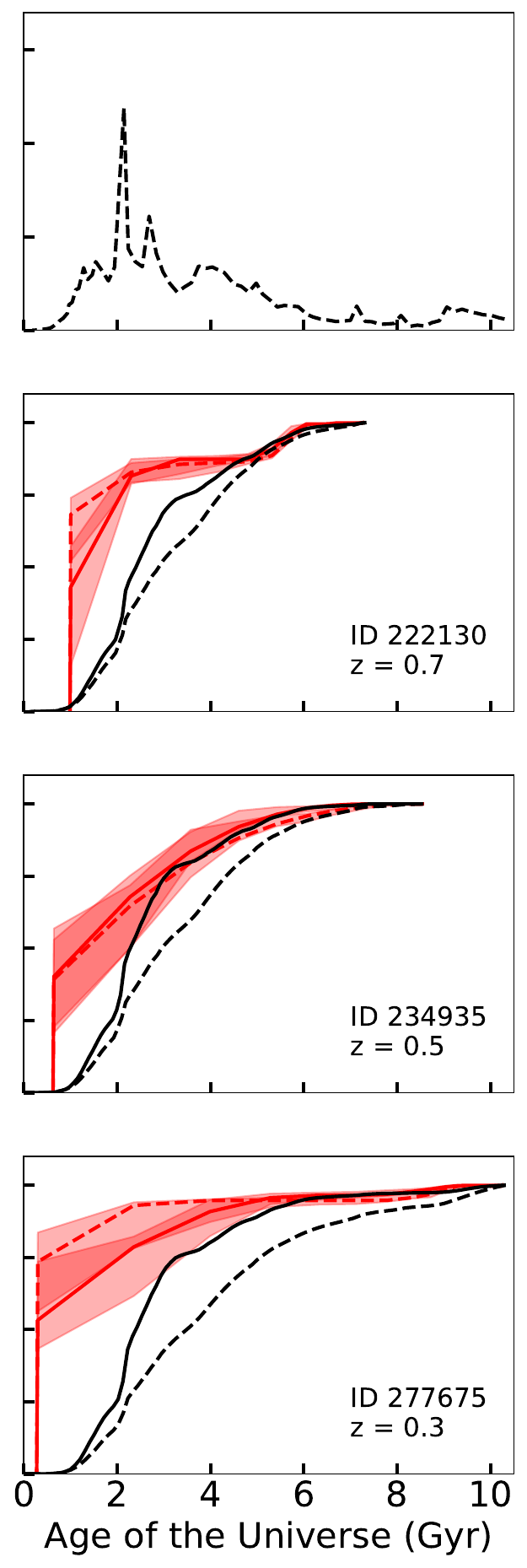}
\includegraphics[width=0.30\textwidth]{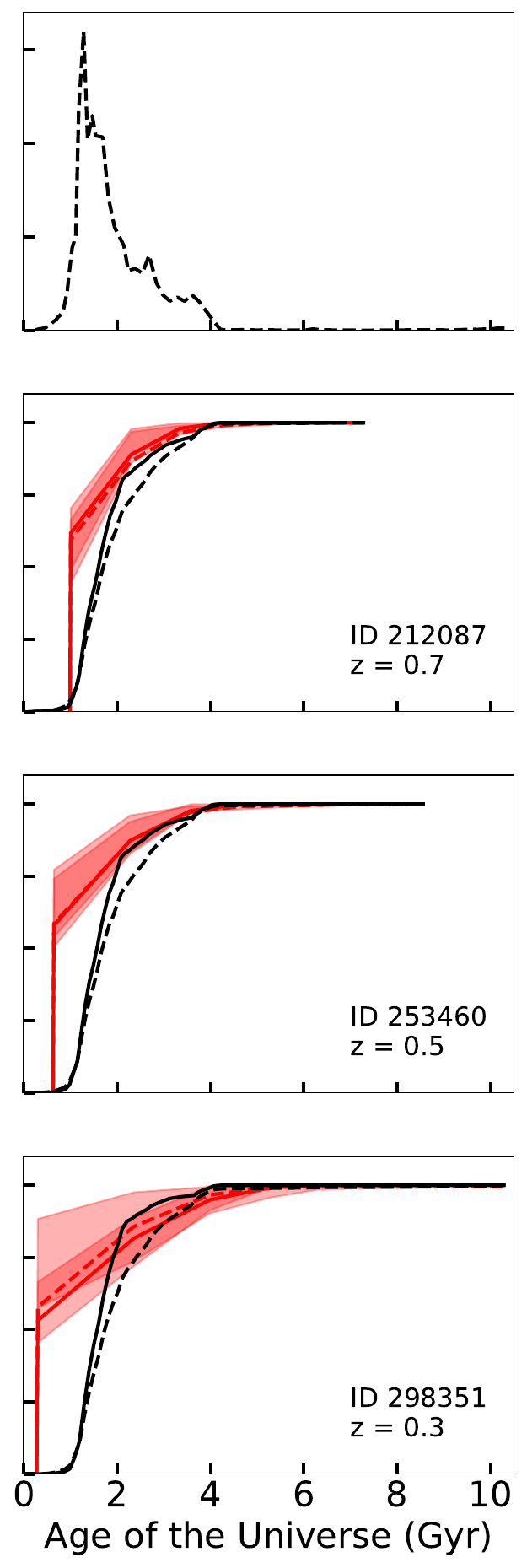}
\caption{
Examples of SFHs (first row) and 
cumulative SFHs of the target galaxies 
at redshift $z=0.7$ (second row), 
$z=0.5$ (third row), and $z=0.3$ (fourth row). 
The SFHs are derived directly from the cosmological simulation and represent the sum of the $SFR$s of all progenitors at each snapshot.
The cumulative SFHs of each galaxy are retrieved by fitting with  \texttt{pPXF} both
the StePS-like fiber spectra (red solid line)
and integrated noiseless spectra (red dashed line). The shaded red regions indicate the 16th–84th percentile range across 100 bootstrap realizations. For comparison, the cumulative SFHs
derived directly from the cosmological simulation
are shown both for the entire galaxy (black dashed line)
and for the central 1.3~arcsec (black solid line).
\label{fig:figure4}}
\end{figure*}

Thanks to TNG50, we have full access to the variety of SFHs and
stellar, gas, and dark matter properties of our sample galaxies.
In Figure~\ref{fig:figure4} (top row), we can appreciate
the complexity of the SFH of the three example galaxies,
which are characterized by multiple episodes of intense star formation.

\begin{table*}[!h]
\caption{Timescale of mass assembly of the example galaxies derived from the central 1.3~arcsec.
\label{tab:table4}}
\centering
\begin{tabular}{c c c c c c c c c}
\hline\hline                 
redshift    & ID     & Class   & $t_{10}^{\rm sim}$    & $t_{50}^{\rm sim}$   & $t_{90}^{\rm sim}$  & $t_{10}^{\rm obs}$    & $t_{50}^{\rm obs}$   & $t_{90}^{\rm obs}$  \\
            &        &         & [Gyr]       & [Gyr]      & [Gyr]     & [Gyr]            & [Gyr] & [Gyr]            \\
(1)         & (2)    & (3)     & (4)         & (5)        & (6)       & (7)              & (8) & (9)              \\
\hline
0.7 & 172231 & SF & 2.00 & 3.86 & 6.13 & 1.22 $\pm$ 0.01 & 2.11 $\pm$ 0.01 & 4.8 $\pm$ 0.3 \\
0.5 & 218606 & SF & 2.03 & 4.05 & 7.44 & 0.63 $\pm$ 0.02 & 2.1 $\pm$ 0.3 & 7.0 $\pm$ 0.1 \\
0.3 & 251598 & GV & 1.94 & 3.63 & 6.56 & 0.29 $\pm$ 0.03 & 3.3 $\pm$ 0.9 & 6.1 $\pm$ 0.5 \\
\hline
0.7 & 222130 & GV & 1.37 & 2.40 & 5.05 & 0.99 $\pm$ 0.01 & 1.24 $\pm$ 0.01 & 5.2 $\pm$ 0.4 \\
0.5 & 234935 & GV & 1.32 & 2.28 & 4.60 & 0.63 $\pm$ 0.01 & 1.2 $\pm$ 0.1 & 4.4 $\pm$ 0.5 \\
0.3 & 277675 & SF & 1.38 & 2.40 & 5.11 & 0.28 $\pm$ 0.01 & 0.30 $\pm$ 0.06 & 3.9 $\pm$ 0.9 \\
\hline
0.7 & 212087 & Q & 1.15 & 1.60 & 2.83 & 0.99 $\pm$ 0.01 & 1.00 $\pm$ 0.01 & 2.4 $\pm$ 0.1 \\
0.5 & 253460 & GV & 1.13 & 1.57 & 2.74 & 0.63 $\pm$ 0.01 & 0.64 $\pm$ 0.01 & 2.6 $\pm$ 0.4 \\
0.3 & 298351 & SF & 1.17 & 1.57 & 2.45 & 0.28 $\pm$ 0.01 & 0.30 $\pm$ 0.01 & 3.4 $\pm$ 0.7 \\
\hline                                  
\end{tabular}
\tablefoot{(1) Redshift. (2) Galaxy ID. (3) Class: SF = star forming, 
GV = green valley, Q = quiescent. 
(4)-(6) Time when the galaxy formed $10\%$, $50\%$, and $90\%$ of its stellar mass, respectively, as derived from TNG50.
(7)-(9) Time when the galaxy formed $10\%$, $50\%$, and $90\%$ of its stellar mass, respectively, as derived with \texttt{pPXF} from noiseless spectra.
}
\end{table*}

To bridge the gap between theoretical and observational descriptions
of the SFH of galaxies, we analyze the noiseless spectra
and compare the 
SFHs of our galaxies retrieved from \texttt{pPXF} with those
inferred from the simulation. To visualize the burstiness of the SFH, we compute the sum of the star formation rates (SFRs) of all progenitors at each snapshot, as shown in the first row of Fig. \ref{fig:figure4}.  In the second to fourth rows, we describe the SFH as the cumulative mass fraction over cosmic time.
From TNG50 (Fig.~\ref{fig:figure4}, dashed lines), 
we derive the cumulative SFH of each galaxy by
integrating across cosmic time
the normalized mass-weighted age
of each star particle either in the 
entire galaxy (dashed black line) or in
the central 1.3~arcsec (dashed red line).
For the noiseless mock spectra (Fig.~\ref{fig:figure4}, 
solid lines), we marginalize over metallicity
the age weights obtained from \texttt{pPXF} and
integrate them across cosmic time. From the recovered cumulative SFHs, 
we derive $t_{10}$, $t_{50}$, and $t_{90}$ 
as the time when the galaxy formed $10\%$, $50\%$ 
and $90\%$ of its stellar mass (Table ~\ref{tab:table4}).

The three progenitor galaxies (and their descendants) 
have very different SFHs:
galaxy ID~212087 formed its stars over a short period of time
($z \sim 4.5$), when the Universe was only 1.3~Gyr old;
ID~222130 had one prominent burst at $z\sim 3$ 
(age of the Universe of 2.1~Gyr) and a residual 
star formation during a few Gyr down to $z\sim0.3$;
ID~172231 shows a very bursty evolution with an extended
period of intense bursts ($SFR > 100~M_{\odot}~{\rm yr}^{-1}$
from $z=2.3$ to $z=1.3$ (age of the Universe 
from 2.8 to 4.8~Gyr) 
and mild residual star formation during a few Gyr
until it quenches at $z \sim 0.3$.

We find that if the galaxy forms at early cosmic times 
in a single burst, such as ID~212087, there is almost
a perfect agreement between the intrinsic and inferred 
cumulative SFH (Fig.~\ref{fig:figure4}, third column).
This could reflect the fact that galaxies forming in one single episode 
have simpler stellar populations \citep[e.g.,][]
{Pappalardo.C:2021}.
In general, if the dominant burst is older than $\sim 5$~Gyr,
the inferred stellar populations are massively old 
\citep[see also][]{GrebolTomas.P:2023}.
The difference between intrinsic and measured 
cumulative SFH is higher for prolonged 
and very bursty SFHs (e.g., ID~172231).
In all cases, we find that the percentage of mass formed in
the first Gyr is greater than 25\%, while the intrinsic values
could be as low as 2-5\%.
Despite this systematic trend, which reflects the fact
that we are not sensitive to the shape of the SFH,
there is better agreement in the timescale when
the galaxies formed almost the totality of its stellar mass,
especially at the highest redshifts,
reflecting the fact that we are very sensitive to the
timescale of the main quenching event.

We find that the example galaxies and their descendants
have similar $t_{90}$ at all redshifts (average difference of $0.4 \pm 0.6$~Gyr).
However, galaxies can show different $t_{10}$ and $t_{50}$,
with inferred values being older
than intrinsic ones (average difference of $0.8 \pm 0.5$~Gyr and $1.2 \pm 0.5$~Gyr).
Again, this reflects the fact that we are overestimating
the fraction of mass forming in the first phases of galaxy evolution,
especially when they have complex SFHs. 
This trend is partially mitigated once comparing
the assigned values during the 
radiative transfer modeling to 
the inferred ones (see Appendix~\ref{sec:appendixC}).
Finally, we calculate the symmetric mean absolute 
percentage error for all galaxies in bins of 1 Gyr 
(lookback time). We notice that there is an almost 
perfect agreement between the predicted and measured 
trends of SFH in a lookback time of 4 Gyr 
(average values < 5\%), except for galaxy ID 172231.

\section{Summary and conclusions \label{sec:section5}}

We use the radiative transfer code \texttt{SKIRT} 
to create mock spectroscopic and imaging datasets 
of galaxies from the TNG50 cosmological simulation.
This dataset represents the ideal solution for 
comparing forthcoming WEAVE-like observations 
with cosmological simulations.
We describe the methodology for creating
the mock datasets and publicly release them
with this work.
Furthermore, we analyze three showcase galaxies
at $z=0.7$ and their descendants at $z=0.5$ and $0.3$,
deriving their mass-weighted age, metallicity,
and star formation history. 

We compare the intrinsic stellar population
properties to those inferred using \texttt{pPXF},
finding that there is an overall good agreement 
in retrieving the cumulative SFH of the showcased galaxies. 
In particular, we are very sensitive to the timescale
when galaxies build up the bulk of their stellar mass,
with a $\lesssim5$\% average difference in the 
cumulative SFH estimations across a lookback time of 4~Gyr
(except for galaxy ID~172231).
Based on the trend derived from the three example galaxies
presented in this work,
selected for the diversity of their SFHs, 
we derive compatible ages but lower
metallicities compared to the intrinsic
age-metallicity distribution retrieved from TNG50.
The analysis of the entire sample of galaxies,
to be presented in a companion paper,
will allow us to statistically explore all 
the possible systematics addressed in this work.
It is worth stressing that, 
while our work relies on the sample selection criteria 
and observational setup of WEAVE-StePS, 
the noiseless datasets can be used
to mimic the observational setup of any facility.
Furthermore, the spectroscopic information can be combined
with the imaging dataset, adding valuable information
about the interplay between the mass build up and the 
morphological transformation of galaxies,
and reducing the age-metallicity 
degeneracy that could hamper our understanding
of mass build-up in the early stages of galaxy evolution.

In conclusion, these mock observations
can be used both for quantifying possible biases
(and the correcting factors) in deriving the SFHs
of galaxies and for properly comparing 
ages, metallicities, and star formation timescales
of observed and simulated galaxies.

\begin{acknowledgements}

We would like to thank the anonymous referee for all the comments that improved the content of the manuscript.
We thank D. Bettoni for the relevant comments during the internal review of the manuscript.
We wish to thank S. Jin for her careful reading of the draft that helped to improve this paper.
We would like to thank M.~Baes for the valuable discussion about \texttt{SKIRT} modeling. 

This project has received funding from the European Union’s Horizon 2020 research and innovation programme under the Marie Skłodowska-Curie Grant Agreement No.101034319 and from the European Union – NextGenerationEU.
The project that gave rise to these results received the support of a fellowship from the “la Caixa” Foundation (ID 100010434). 
The fellowship code is LCF/BQ/PR24/12050015. LC acknowledges support from grants PID2022-139567NB-I00 and PIB2021-127718NB-I00 funded by the Spanish Ministry of Science and Innovation/State Agency of Research  MCIN/AEI/10.13039/501100011033 and by “ERDF A way of making Europe”. 
AFM acknowledges support from RYC2021-031099-I and PID2021-123313NA-I00 of MICIN/AEI/10.13039/501100011033/FEDER, UE, NextGenerationEU/PRT. 
EMC and AP are supported by the Istituto Nazionale di Astrofisica (INAF) through the grant Progetto di Ricerca di Interesse Nazionale (PRIN) 2022 2022383WFT "SUNRISE" (CUP C53D23000850006) and Padua University with the grants Dotazione Ordinaria Ricerca (DOR) 2021-2023.
A.I., F.D., M.L. and S.Z. acknowledge financial support from INAF Mainstream grant 2019 WEAVE StePS 1.05.01.86.16 and INAF Large Grant 2022 WEAVE StePS 1.05.12.01.11. 
PSB acknowledges support from Grant  PID2022-138855NB-C31 funded by MICIU/AEI/10.13039/501100011033 and by ERDF/EU.
R.R. acknowledges financial support grants through INAF-WEAVE StePS founds and through PRIN-MIUR 2020SKSTHZ.
C.P.H. acknowledges support from ANID through Fondecyt Regular project number 1252233.
Funding for the WEAVE facility has been provided by UKRI STFC, the University of Oxford, NOVA, NWO, Instituto de Astrofísica de Canarias (IAC), the Isaac Newton Group partners (STFC, NWO, and Spain, led by the IAC), INAF, CNRS-INSU, the Observatoire de Paris, Région Île-de-France, CONACYT through INAOE, the Ministry of Education, Science and Sports of the Republic of Lithuania, Konkoly Observatory (CSFK), Max-Planck-Institut für Astronomie (MPIA Heidelberg), Lund University, the Leibniz Institute for Astrophysics Potsdam (AIP), the Swedish Research Council, the European Commission, and the University of Pennsylvania.  The WEAVE Survey Consortium consists of the ING, its three partners, represented by UKRI STFC, NWO, and the IAC, NOVA, INAF, GEPI, INAOE, Vilnius University, FTMC – Center for Physical Sciences and Technology (Vilnius), and individual WEAVE Participants. Please see the relevant footnotes for the WEAVE website\footnote{\url{https://weave-project.atlassian.net/wiki/display/WEAVE}} and for the full list of granting agencies and grants supporting WEAVE\footnote{\url{https://weave-project.atlassian.net/wiki/display/WEAVE/WEAVE+Acknowledgements}}.

\end{acknowledgements}
    
\bibliographystyle{aa}
\bibliography{ikhsanova+25}

\begin{appendix} 

\onecolumn 

\section{Example of mock imaging dataset \label{sec:appendixA}}

\begin{figure*}[!h] 
\centering
\includegraphics[width=0.85\textwidth]{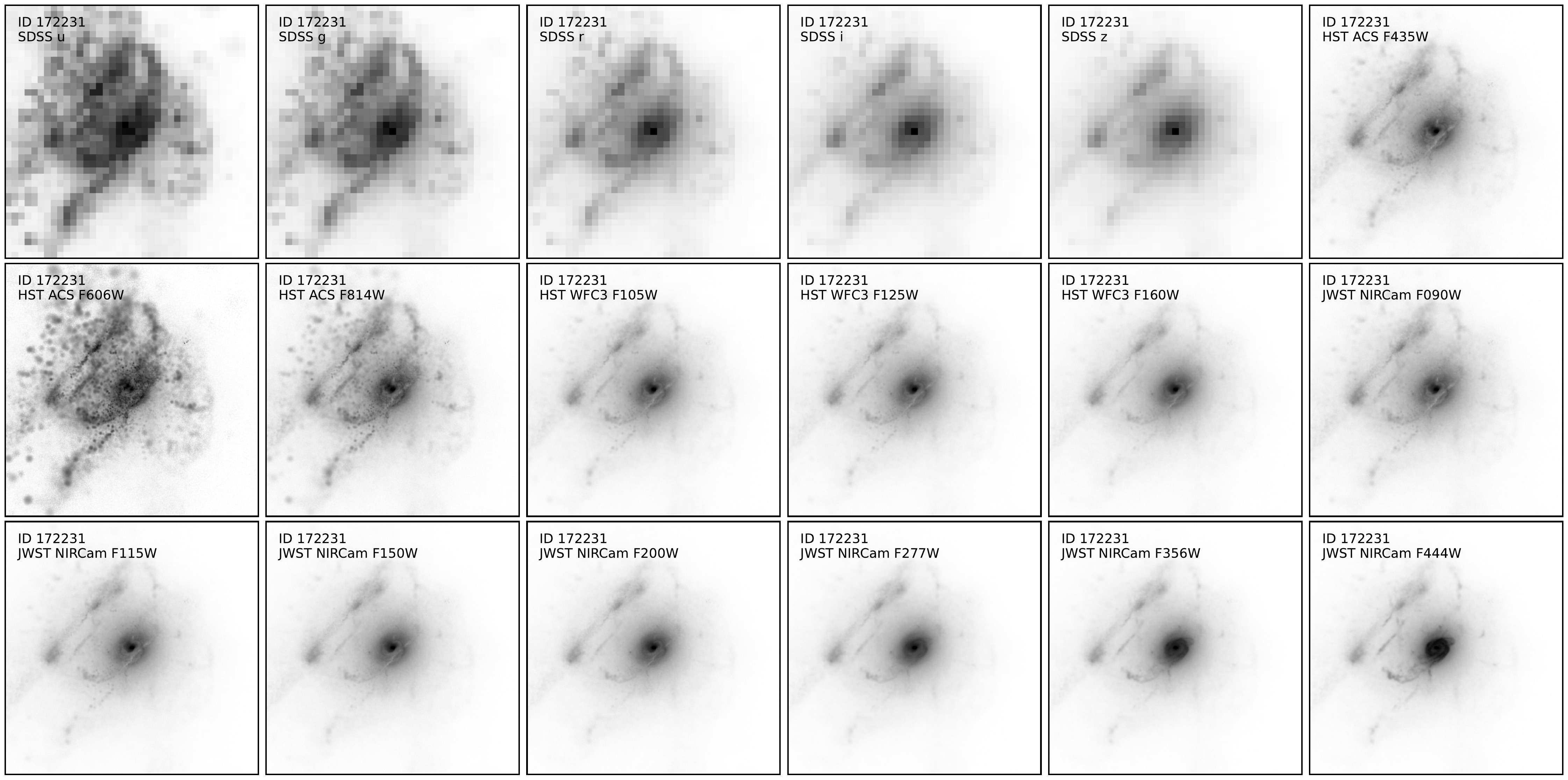}
\caption{Example of the multi-wavelength imaging datasets 
for galaxy ID~172231.
\label{fig:figureA}}
\end{figure*}

As a complementary dataset to the spectroscopic one, 
for each galaxy we create 18 mock images 
covering the optical to NIR regime
with different spatial resolutions, as detailed 
in Sect.~\ref{sec:section3.3}.
In Fig.~\ref{fig:figureA}, we present an example 
of the multi-wavelength imaging dataset for galaxy ID~172231.

\FloatBarrier

\section{Example of full-spectral fitting \label{sec:appendixB}}

\begin{figure*}[!h]     
\centering
\includegraphics[width=0.45\textwidth]{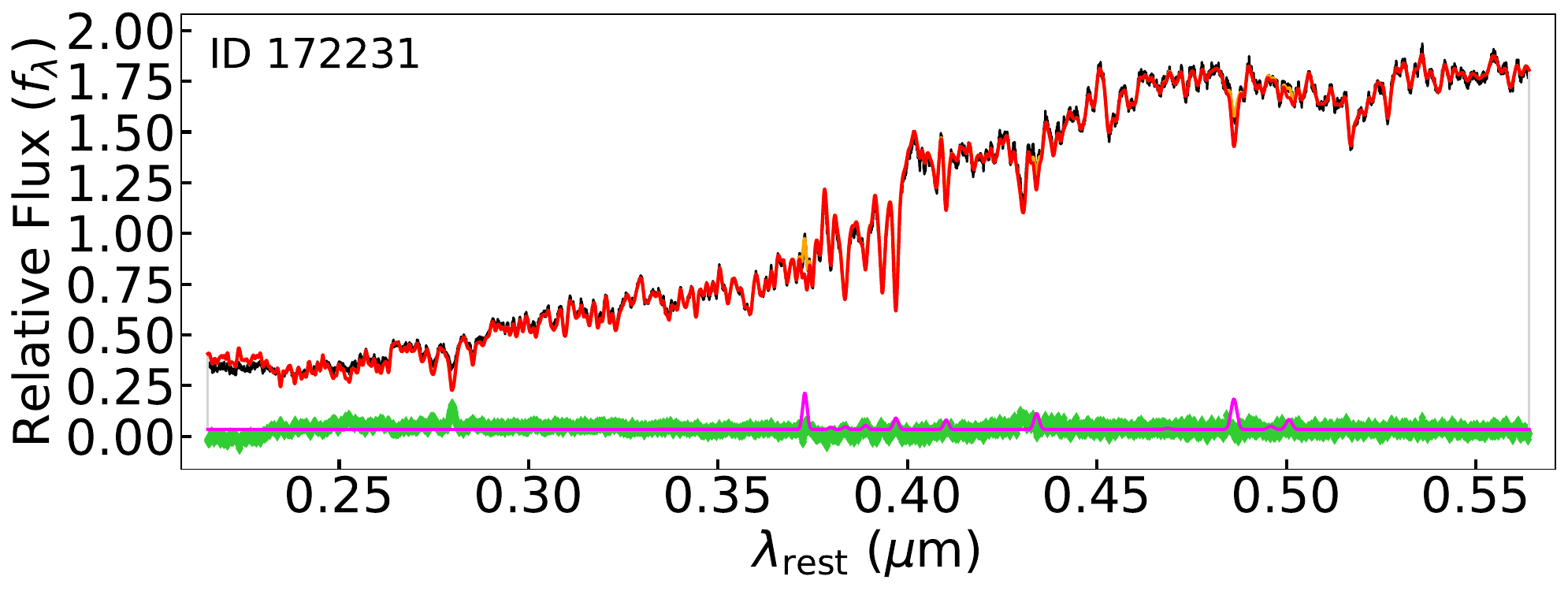}
\includegraphics[width=0.45\textwidth]{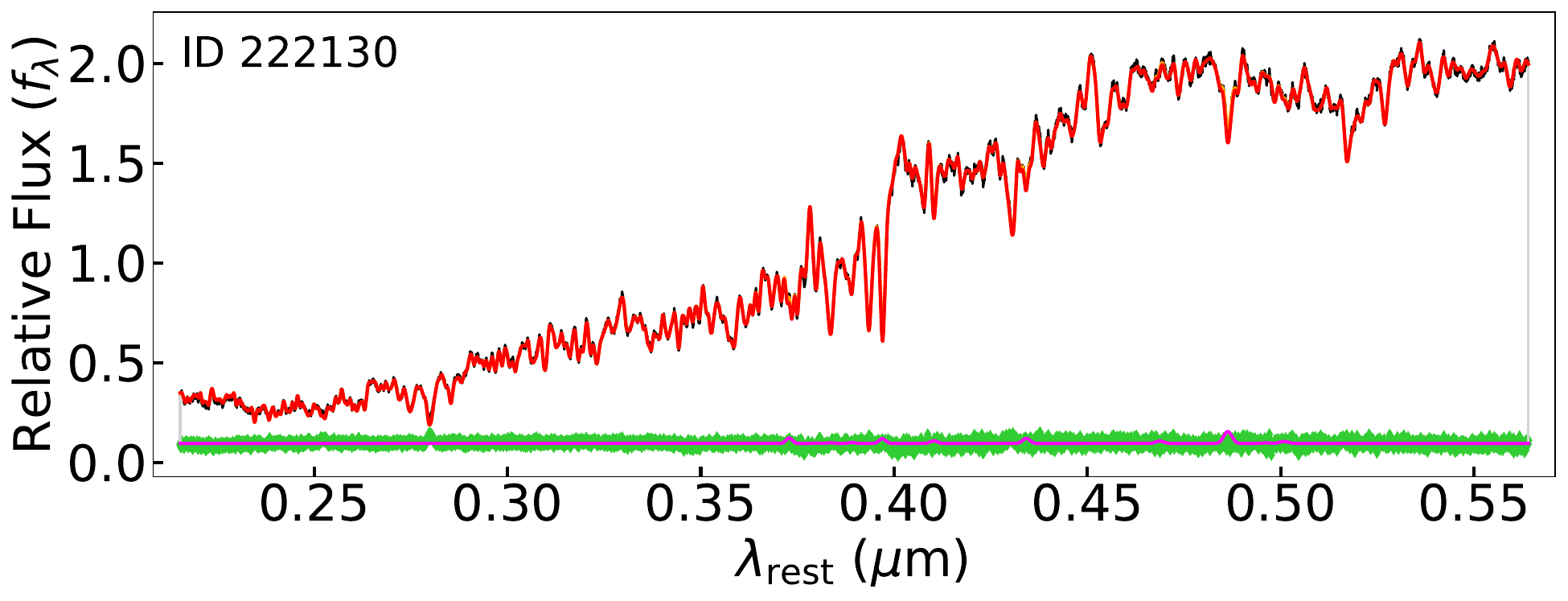}
\includegraphics[width=0.45\textwidth]{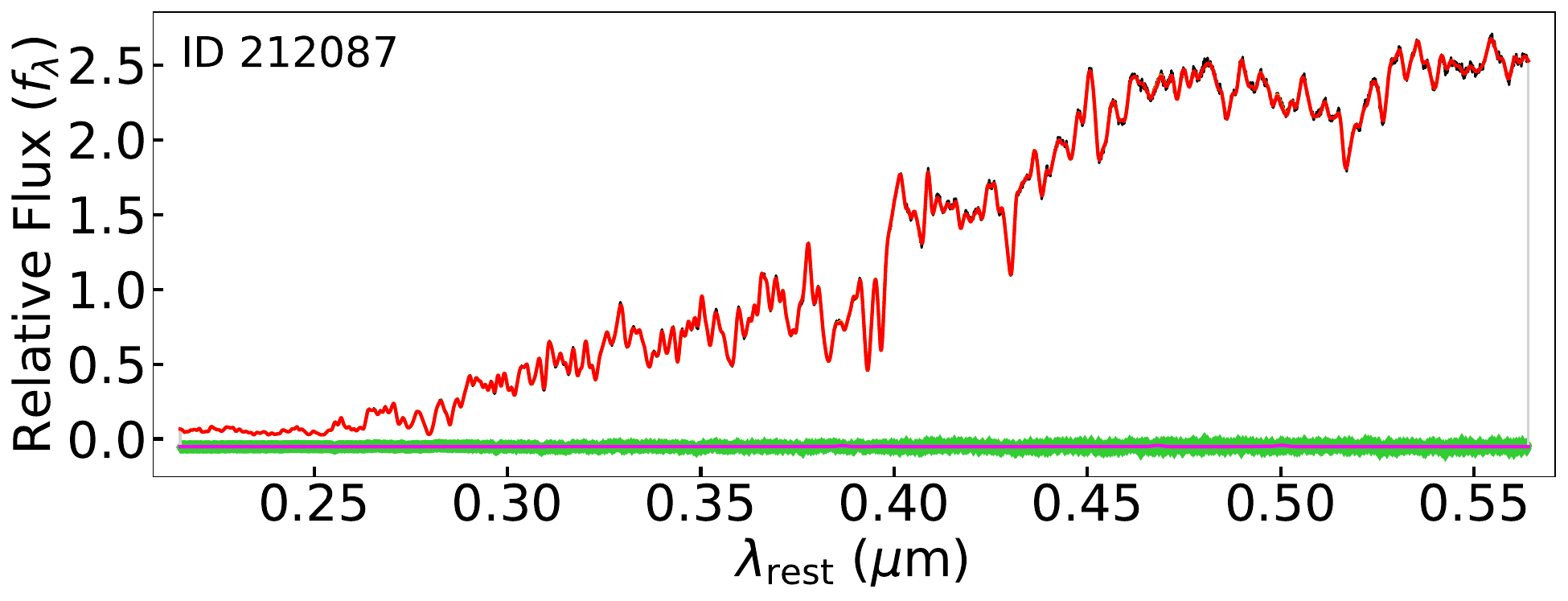}
\caption{
Example of full spectral fitting performed with \texttt{pPXF}
for three galaxies at redshift $z = 0.7$.
The rest-frame noiseless spectra (black solid lines) 
are plotted with the best-fitting model (orange lines).
Each emission line is modelled 
with one kinematic component.
The best-fitting stellar spectrum alone and 
the gas emissions alone
are shown with red and magenta
solid lines, respectively.
The residuals (green diamonds; arbitrarily offset) 
are calculated by subtracting the model 
from the observed spectrum. 
}
\label{fig:figureB}
\end{figure*}

In Fig.~\ref{fig:figureB}, we report 
an example of the spectral fitting results
performed with \texttt{pPXF} for the three progenitor
galaxies at $z=0.7$.

\FloatBarrier

\section{Possible systematics in the radiative transfer modeling \label{sec:appendixC}}

In this Appendix, we address the main systematic 
introduced in the radiative transfer modeling,
which has to be kept in mind for a proper comparison 
between observations and simulations.

In Sect.~\ref{sec:section3.1}, we model each star
particle older than 10~Myr with a SSP from the BPASS
library. However, once assigning the correspondent 
SSP within \texttt{SKIRT}, 
we are limited by the BPASS metallicity grid,
which, in particular, does not reproduce the highest metallicities in TNG50. 
Furthermore, we analyzed the noiseless spectra 
considering models with stellar ages $>50$~Myr
using the same procedure as in \citet{Cappellari.M:2023}. 
We tested whether this last choice would not affect our results by repeating the analysis including stellar ages down to 1~Myr and found no systematics
in the retrieved stellar population properties.

\begin{figure*}[!h] 
\centering
\includegraphics[width=6cm]{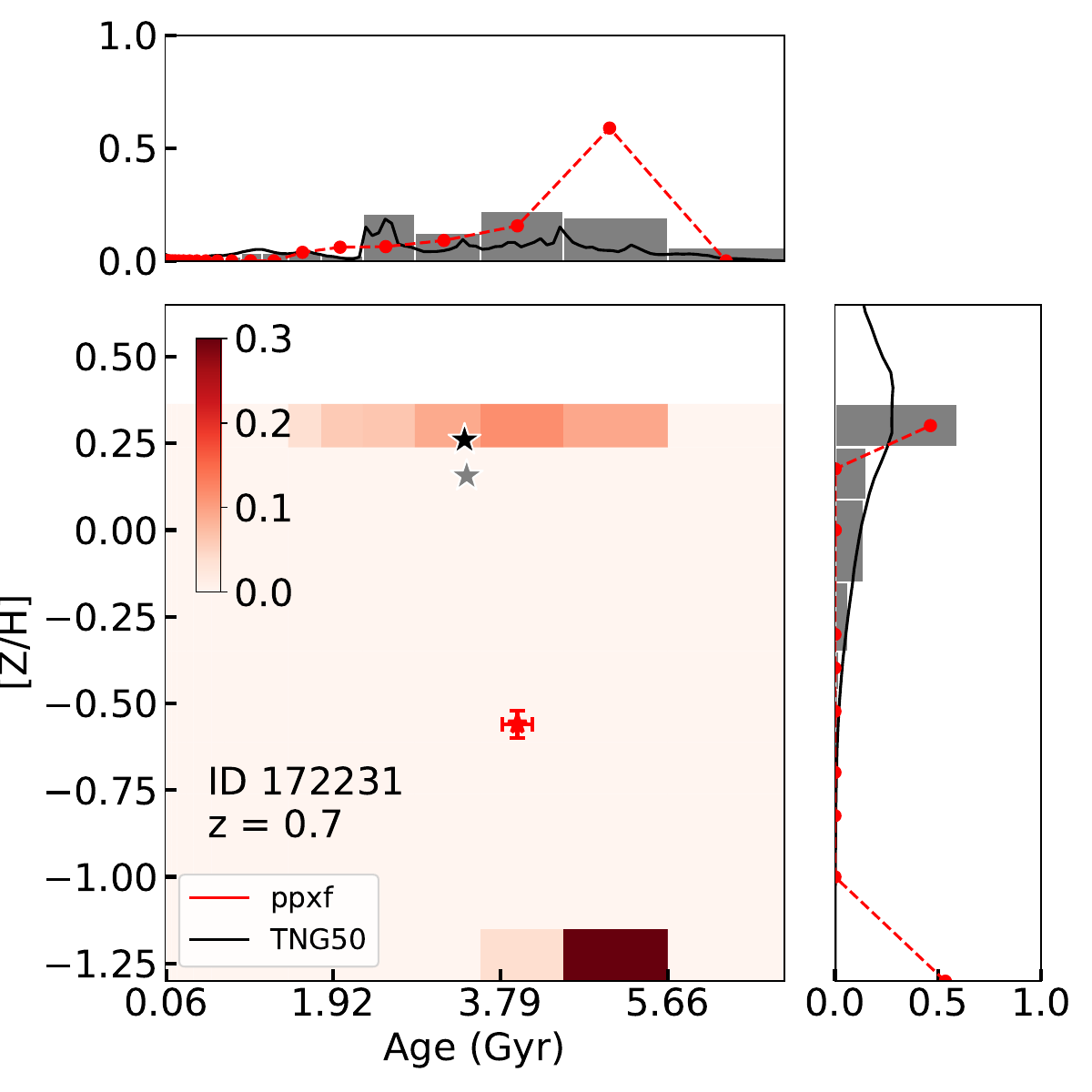}
\includegraphics[width=6cm]{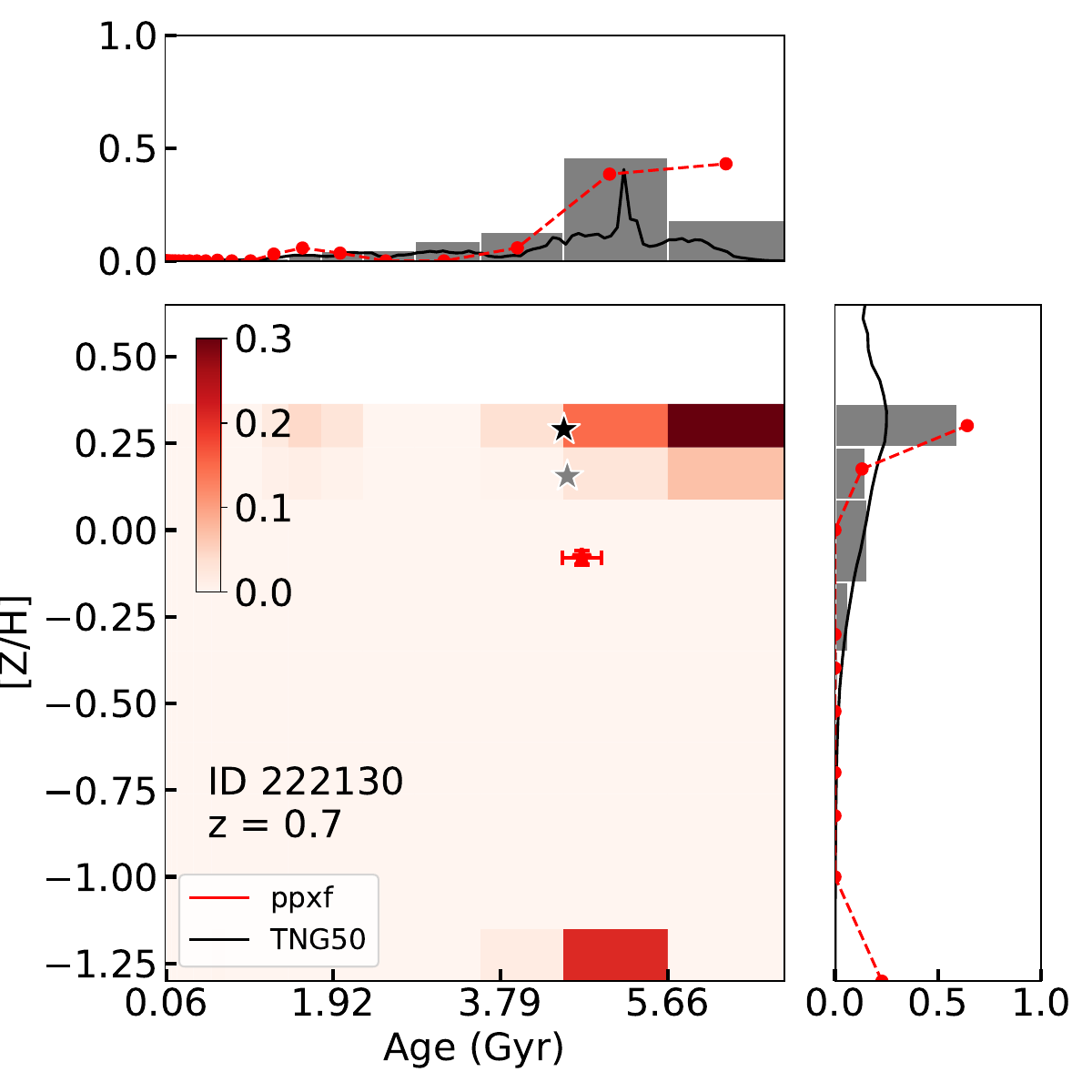}
\includegraphics[width=6cm]{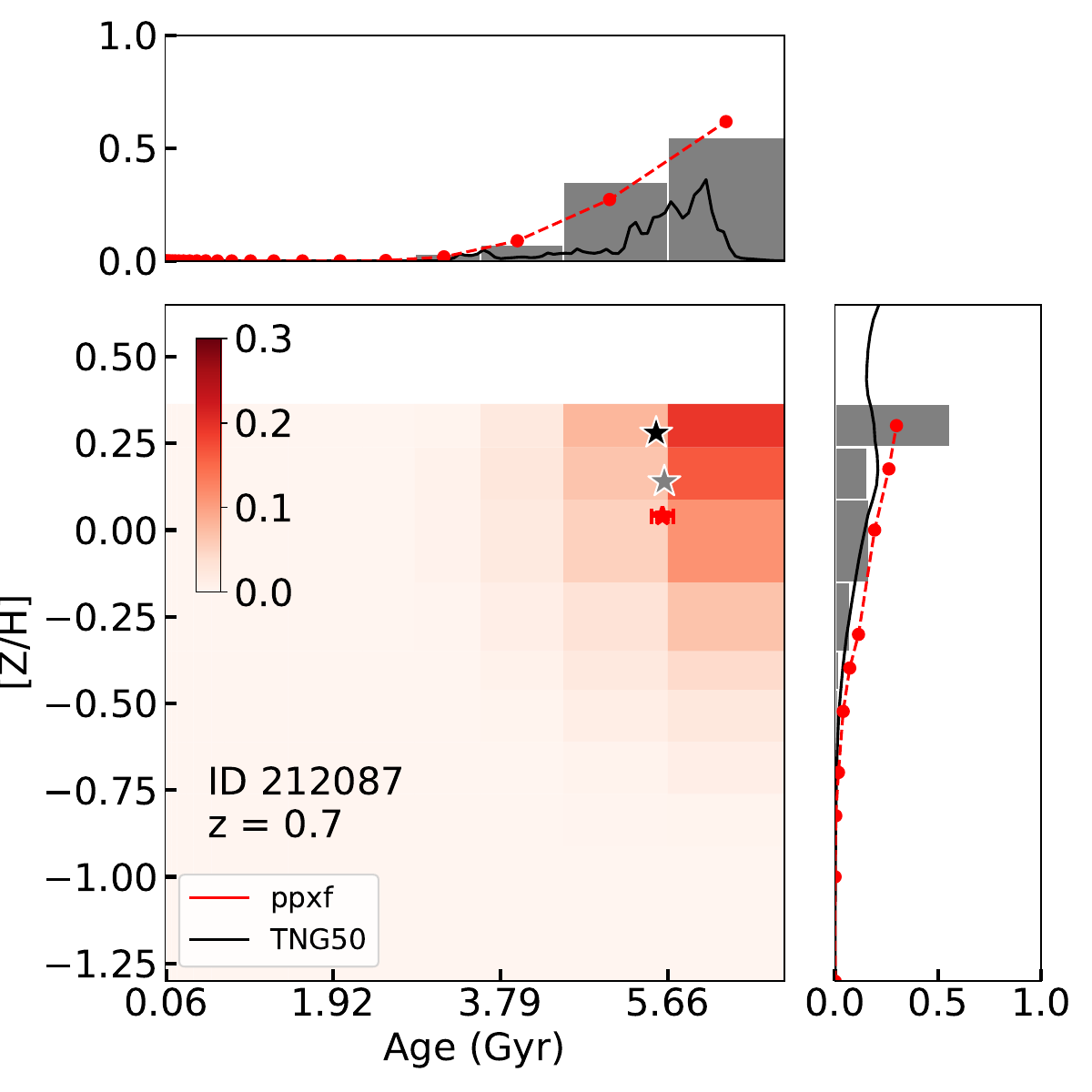}

\includegraphics[width=6cm]{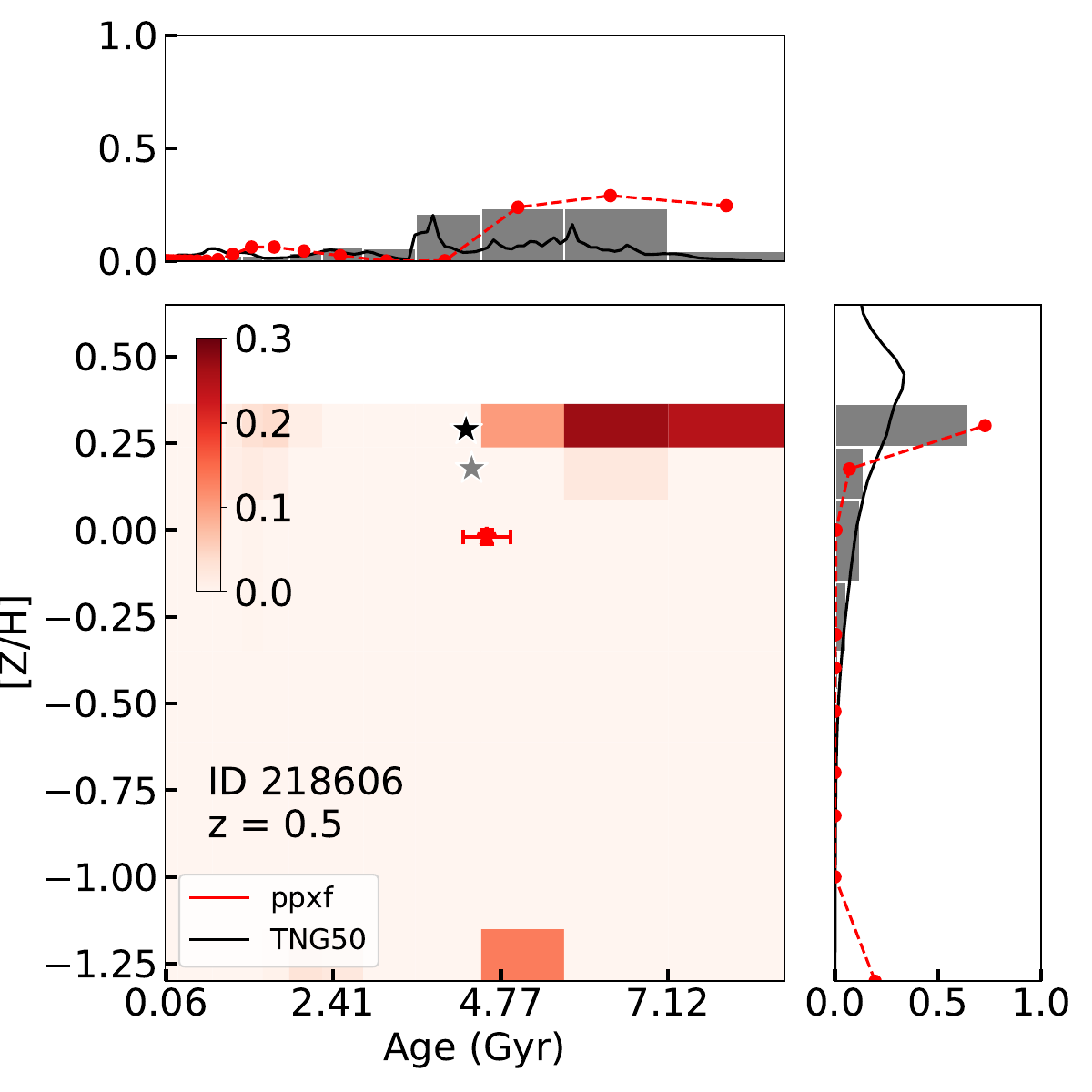}
\includegraphics[width=6cm]{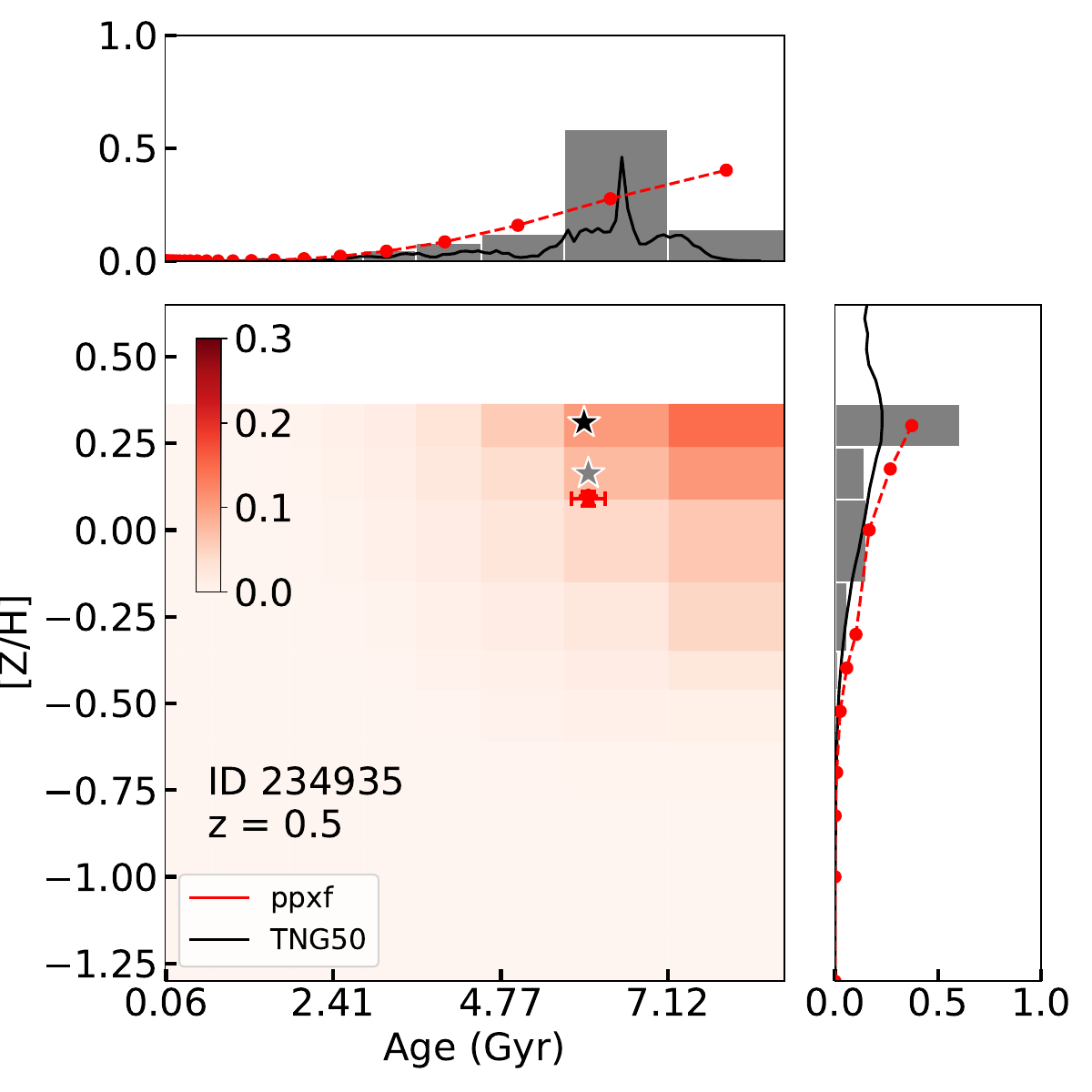}
\includegraphics[width=6cm]{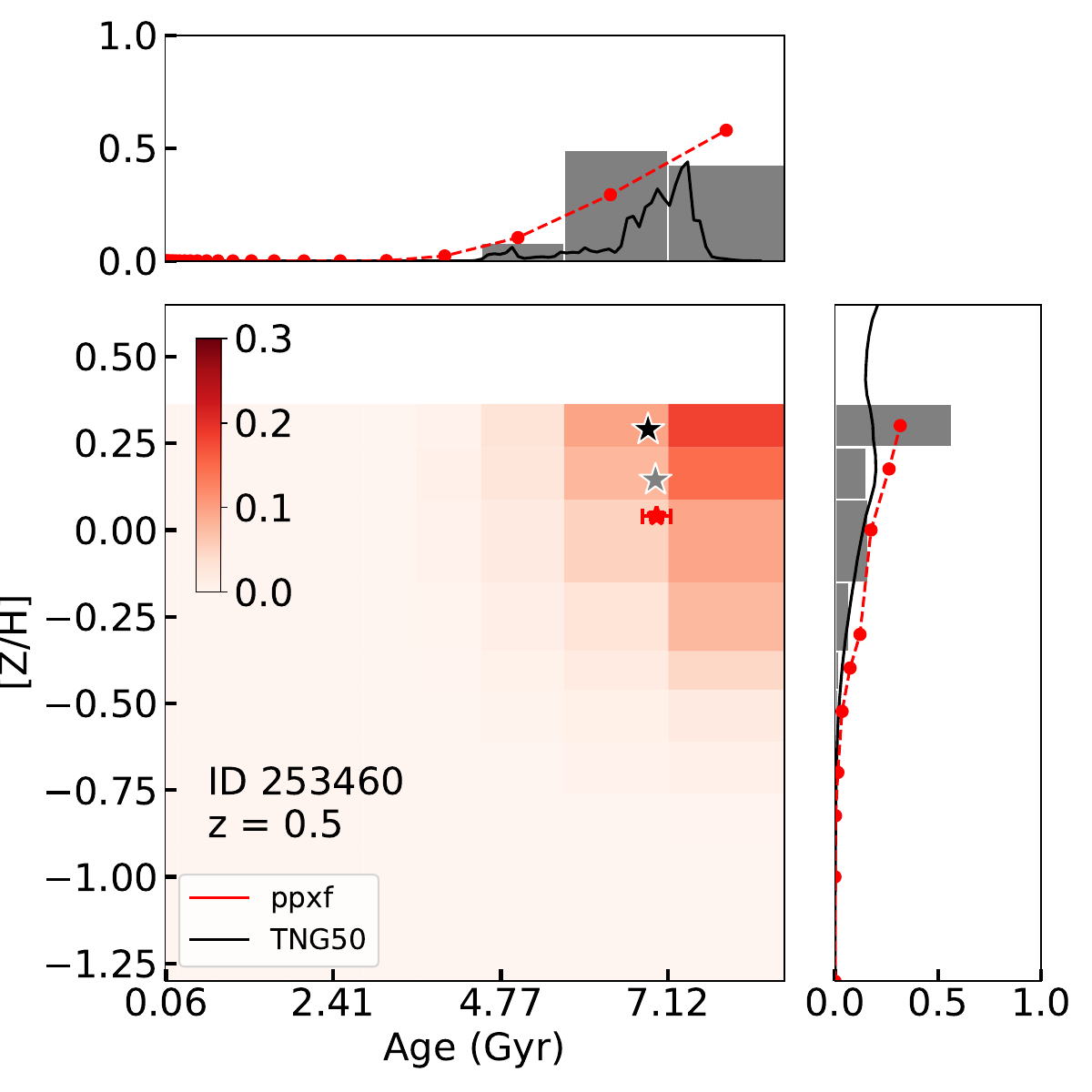}

\includegraphics[width=6cm]{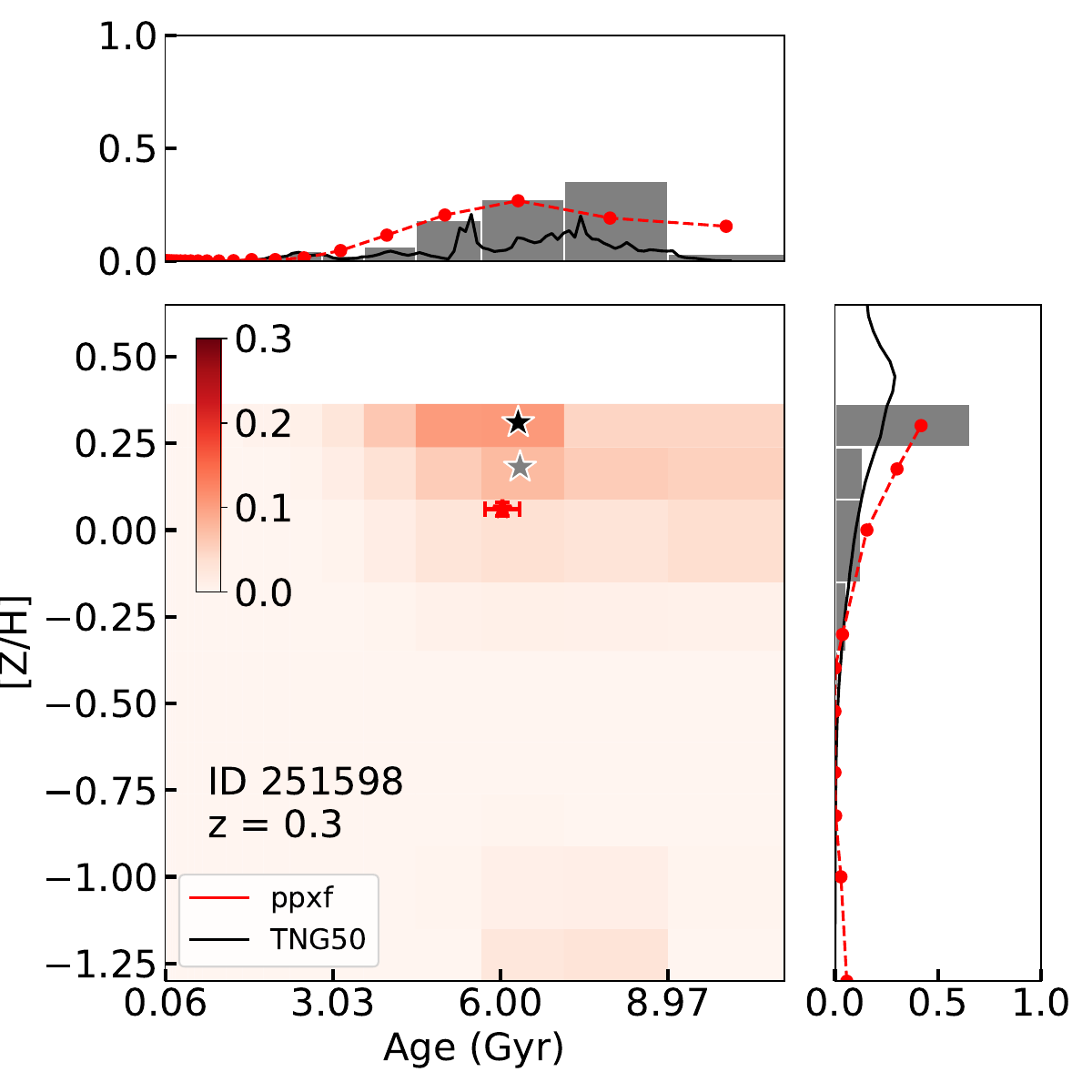}
\includegraphics[width=6cm]{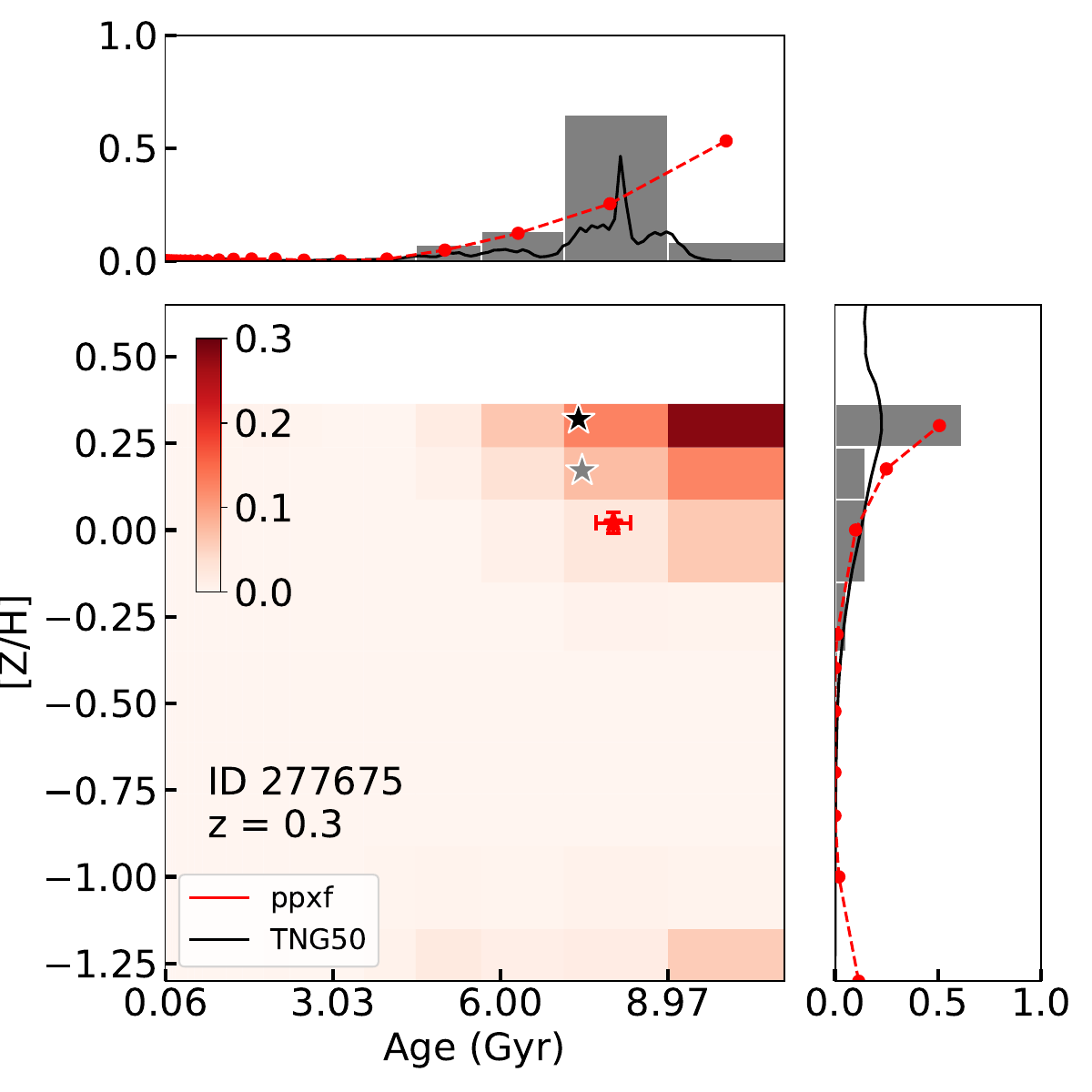}
\includegraphics[width=6cm]{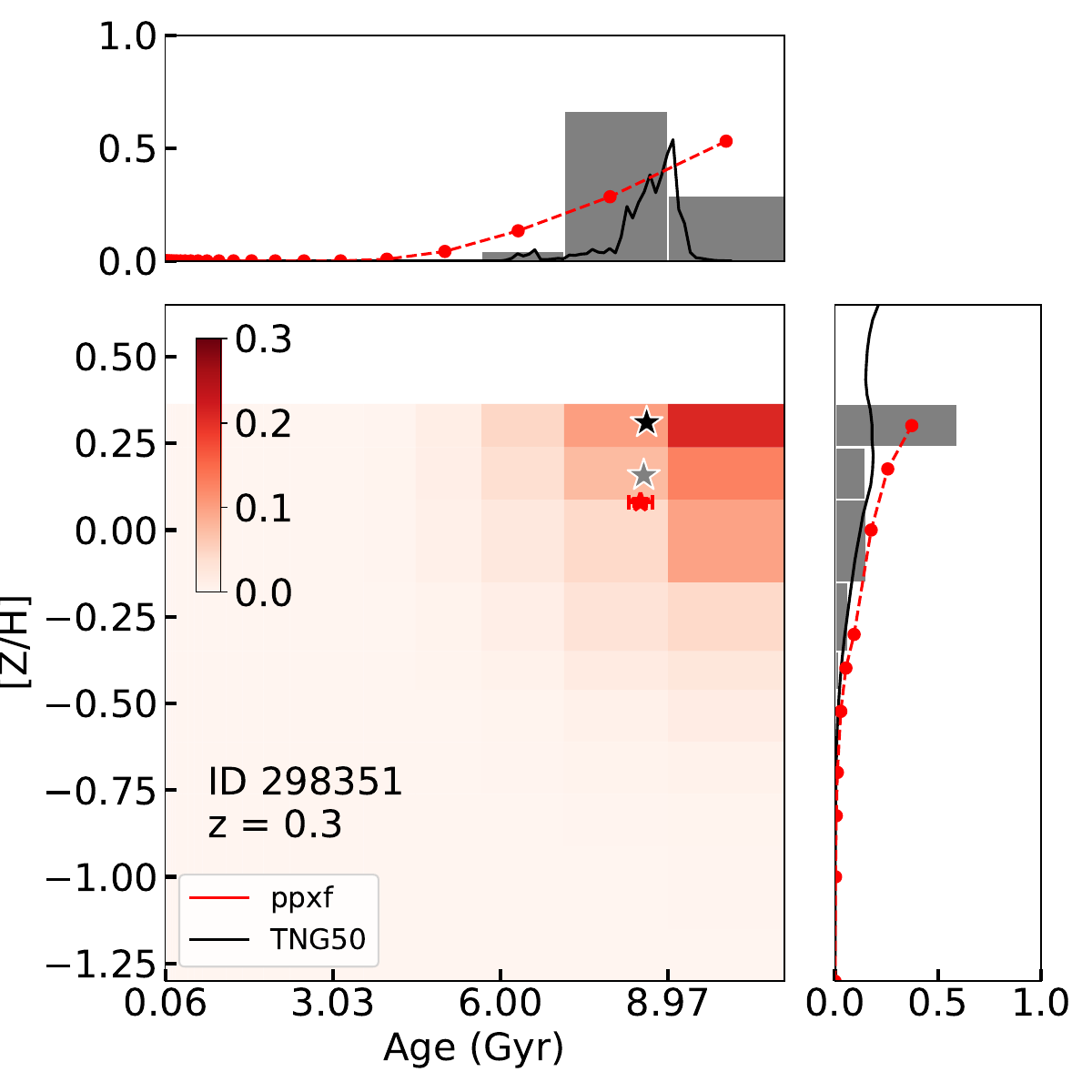}

\caption{Same as Fig.~\ref{fig:figure3}, 
but gray stars representing assigned values 
in \texttt{SKIRT}.
\label{fig:figureC1}}
\end{figure*}

\begin{figure*}[!h]     
\centering
\includegraphics[width=0.365\textwidth]{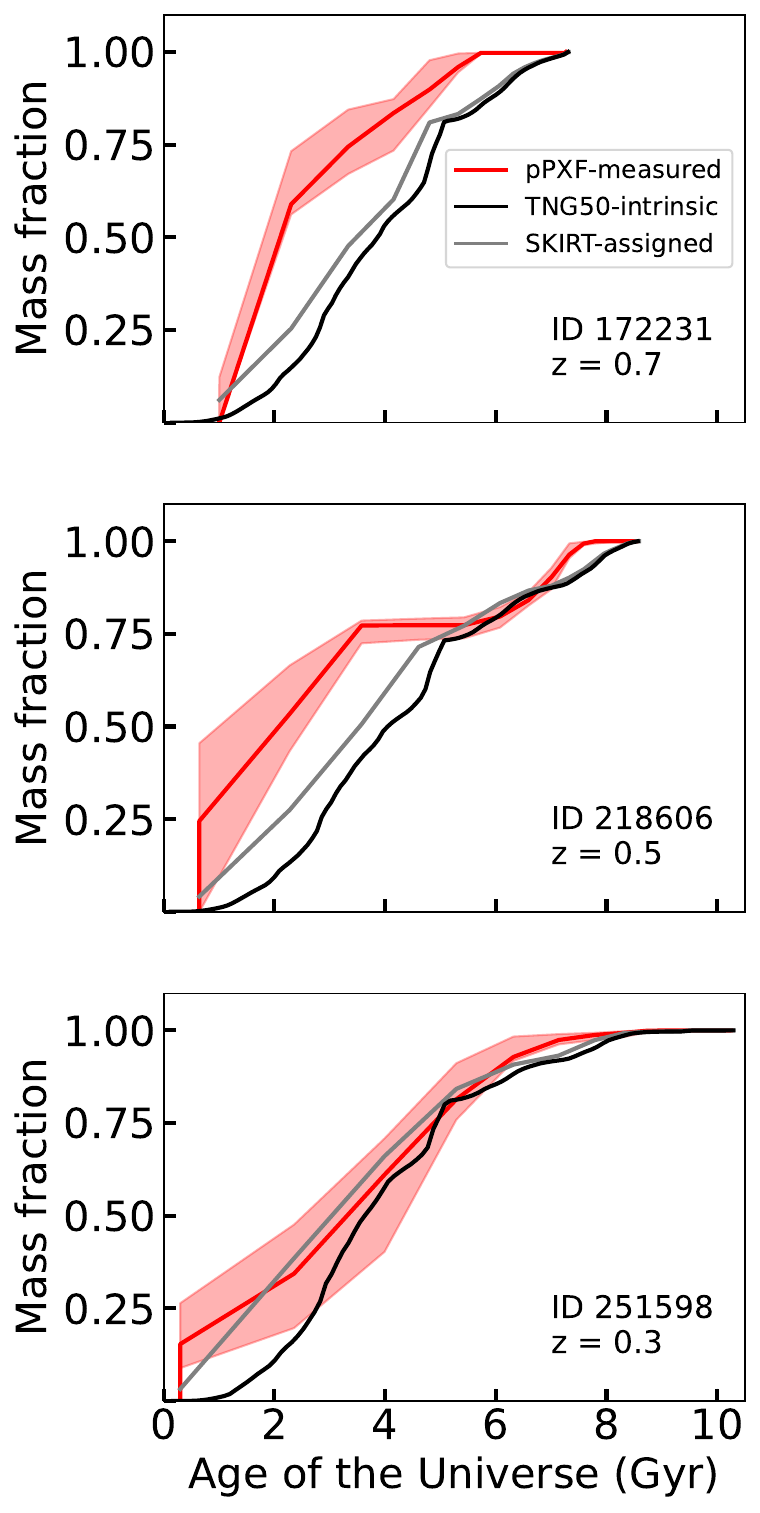}
\includegraphics[width=0.30\textwidth]{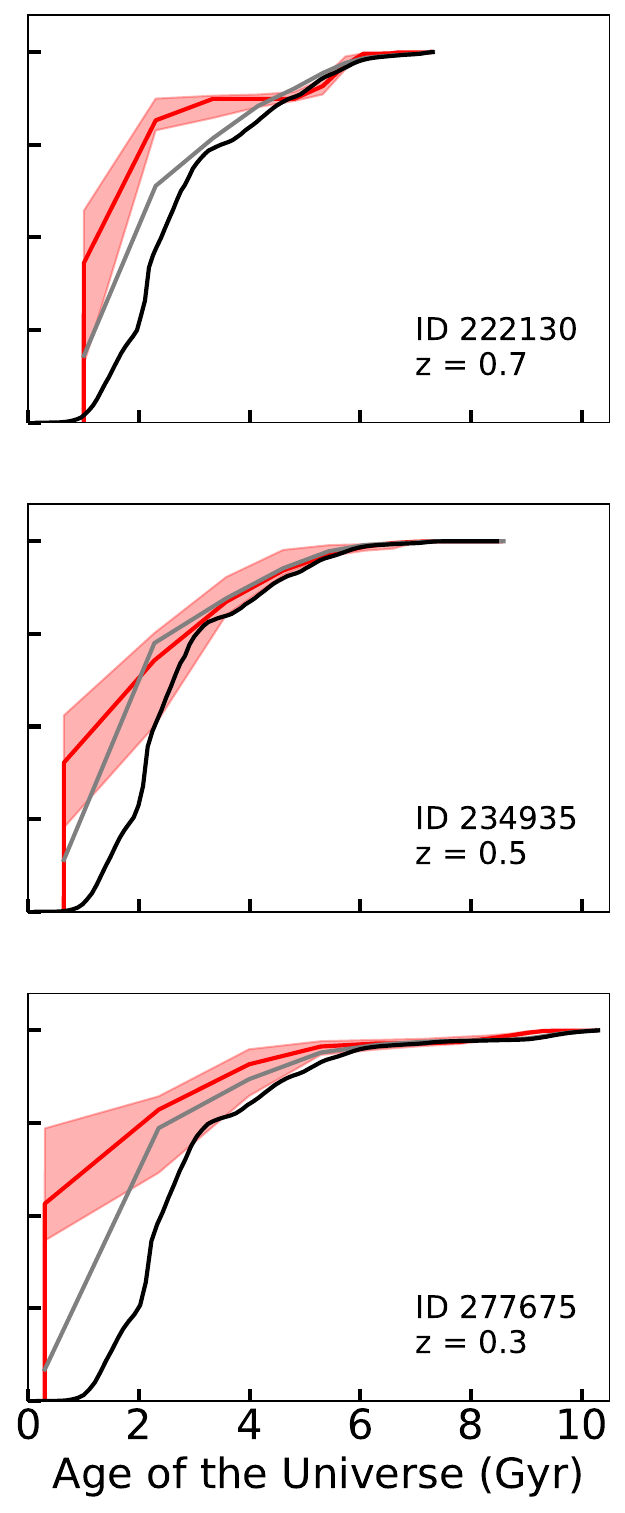}
\includegraphics[width=0.30\textwidth]{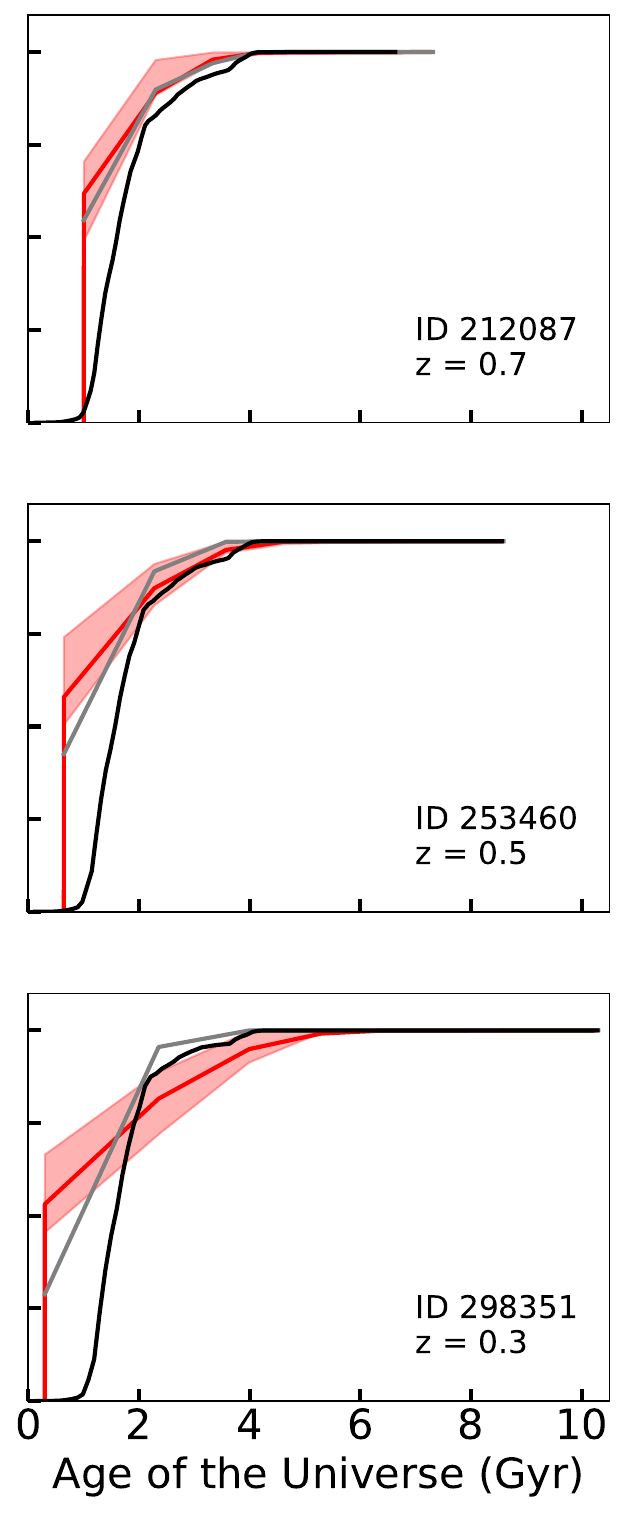}
\caption{
Same as Fig.~\ref{fig:figure4}, 
but gray solid lines representing assigned values 
in radiative transfer modeling (i.e., \texttt{SKIRT}) 
for the fiber spectra.
\label{fig:figureC2}}
\end{figure*}

In Figs.~\ref{fig:figureC1} and \ref{fig:figureC2},
we show the same analysis described 
in Sects.~\ref{sec:section4.1.2} and \ref{sec:section4.2}
but considering also the assigned values 
together with the intrinsic ones.
In this case, the comparison between the
assigned values and those retrieved with \texttt{pPXF}
show a slightly better agreement in the mass-weighted ages,
with an average difference of $0.1 \pm 0.3$~Gyr.
Overall, the assigned SFHs are only slightly 
closer to the inferred ones, 
highlighting that part of the systematics
are not due to the radiative transfer calculations,
especially in the last $\sim4$~Gyr of galaxy evolution
(see also Table~\ref{tab:tablec2}).

\begin{table*}[!h] 
\caption{Mass-weighted ages and  metallicities of the central part of the example galaxies extended with assigned values. 
\label{tab:tablec1}} 
\centering
\begin{tabular}{c c c c c c c c c  } 
\hline\hline 
$z$    & ID     & Class    & age$_{\rm sim}^{\rm intrinsic}$  & age$_{\rm sim}^{\rm assigned}$     & age$_{\rm obs}^{\rm fiber}$   & [Z/H]$_{\rm sim}^{\rm intrinsic}$  & [Z/H]$_{\rm sim}^{\rm assigned}$   & [Z/H]$_{\rm obs}^{\rm fiber}$   \\
            &        &                         & [Gyr]                           & [Gyr]            & [Gyr] & [dex] & [dex]  & [dex]       \\
(1) & (2) & (3)  & (4) & (5)  & (6)  & (7)  & (8)  & (9)       \\
\hline
0.7 & 172231 & SF & 3.39 & 3.42 & 4.0 $\pm$ 0.2 & 0.26 & 0.15 & $-0.56$ $\pm$ 0.04 \\
0.5 & 218606 & SF & 4.28 & 4.36 & 4.6 $\pm$ 0.3 & 0.29 & 0.18 & $-0.02$ $\pm$ 0.02 \\
0.3 & 251598 & GV & 6.31 & 6.35 & 6.0 $\pm$ 0.3 & 0.31 & 0.18 & 0.06 $\pm$ 0.03 \\
\hline
0.7 & 222130 & GV & 4.50 & 4.55 & 4.7 $\pm$ 0.2 & 0.29 & 0.15 & $-0.08$ $\pm$ 0.02 \\
0.5 & 234935 & GV & 5.94 & 6.02 & 6.0 $\pm$ 0.3 & 0.31 & 0.16 & 0.09 $\pm$ 0.02 \\
0.3 & 277675 & SF & 7.38 & 7.46 & 8.0 $\pm$ 0.4 & 0.32 & 0.17 & 0.02 $\pm$ 0.03 \\
\hline
0.7 & 212087 & Q & 5.53 & 5.67 & 5.6 $\pm$ 0.1 & 0.28 & 0.14 & 0.04 $\pm$ 0.01 \\
0.5 & 253460 & GV & 6.84 & 6.99 & 7.0 $\pm$ 0.2 & 0.29 & 0.14 & 0.04 $\pm$ 0.01 \\
0.3 & 298351 & SF & 8.59 & 8.57 & 8.5 $\pm$ 0.2 & 0.31 & 0.16 & 0.08 $\pm$ 0.01 \\
\hline                                  
\end{tabular}
\tablefoot{(1) Redshift. (2) Galaxy ID. (3) Class: SF = star forming, 
GV = green valley, Q = quiescent. 
(4) Mass-weighted age of the central 1.3~arcsec derived from TNG50.
(5) Mass-weighted age of the central 1.3~arcsec assigned by \texttt{SKIRT}.
(6) Mass-weighted age of the central 1.3~arcsec derived with \texttt{pPXF} from noiseless spectra.
(7) Mass-weighted metallicity of the central 1.3~arcsec derived from TNG50.
(8) Mass-weighted metallicity of the central 1.3~arcsec assigned by \texttt{SKIRT}.
(9) Mass-weighted metallicity of the central 1.3~arcsec derived with \texttt{pPXF} from noiseless spectra.}
\end{table*}

\begin{table*}[!h] 
\caption{Same as Table~\ref{tab:table4} but extended with assigned values. 
\label{tab:tablec2}}
\centering
\begin{tabular}{c c c c c c c c c c c c}
\hline\hline                 
redshift    & ID     & Class   & $t_{10}^{\rm sim}$    & $t_{50}^{\rm sim}$   & $t_{90}^{\rm sim}$ & $t_{10}^{\rm assigned}$    & $t_{50}^{\rm assigned}$   & $t_{90}^{\rm assigned}$ & $t_{10}^{\rm obs}$    & $t_{50}^{\rm obs}$   & $t_{90}^{\rm obs}$  \\
            &        &         & [Gyr]       & [Gyr]      & [Gyr]     & [Gyr]& [Gyr]      & [Gyr]     & [Gyr]            & [Gyr] & [Gyr]            \\
(1)         & (2)    & (3)     & (4)         & (5)        & (6)       & (7)              & (8) & (9) & (10)              & (11) & (12)              \\
\hline
0.7 & 172231 & SF & 2.00 & 3.86 & 6.13 & 1.23 & 3.45 & 5.94 & 1.22 $\pm$ 0.01 & 2.11 $\pm$ 0.01 & 4.8 $\pm$ 0.3 \\
0.5 & 218606 & SF & 2.03 & 4.05 & 7.44 & 1.02 & 3.52 & 7.28 & 0.63 $\pm$ 0.02 & 2.1 $\pm$ 0.3 & 7.0 $\pm$ 0.1 \\
0.3 & 251598 & GV & 1.94 & 3.63 & 6.56 & 0.65 & 2.97 & 6.18 & 0.29 $\pm$ 0.03 & 3.3 $\pm$ 0.9 & 6.1 $\pm$ 0.5 \\
\hline
0.7 & 222130 & GV & 1.37 & 2.40 & 5.05 & 1.00 & 1.89 & 4.76 & 0.99 $\pm$ 0.01 & 1.24 $\pm$ 0.01 & 5.2 $\pm$ 0.4 \\
0.5 & 234935 & GV & 1.32 & 2.28 & 4.60 & 0.64 & 1.61 & 4.23 & 0.63 $\pm$ 0.01 & 1.2 $\pm$ 0.1 & 4.4 $\pm$ 0.5 \\
0.3 & 277675 & SF & 1.38 & 2.40 & 5.11 & 0.30 & 1.57 & 4.51 & 0.28 $\pm$ 0.01 & 0.30 $\pm$ 0.06 & 3.9 $\pm$ 0.9 \\
\hline
0.7 & 212087 & Q & 1.15 & 1.60 & 2.83 & 0.99 & 1.00 & 2.29 & 0.99 $\pm$ 0.01 & 1.00 $\pm$ 0.01 & 2.4 $\pm$ 0.1 \\
0.5 & 253460 & GV & 1.13 & 1.57 & 2.74 & 0.63 & 0.81 & 2.20 & 0.63 $\pm$ 0.01 & 0.64 $\pm$ 0.01 & 2.6 $\pm$ 0.4 \\
0.3 & 298351 & SF & 1.17 & 1.57 & 2.45 & 0.28 & 0.81 & 2.17 & 0.28 $\pm$ 0.01 & 0.30 $\pm$ 0.01 & 3.4 $\pm$ 0.7 \\
\hline                                  
\end{tabular}
\tablefoot{(1) Redshift. (2) Galaxy ID. (3) Class: SF = star forming, 
GV = green valley, Q = quiescent. 
(4)-(6) Time when the galaxy formed $10\%$, $50\%$, and $90\%$ of its stellar mass, as derived from TNG50.
(7)-(9) Time when the galaxy formed $10\%$, $50\%$, and $90\%$ of its stellar mass assigned during radiative transfer calculations.
(10)-(12) Time when the galaxy formed $10\%$, $50\%$, and $90\%$ of its stellar mass, as derived with \texttt{pPXF} from noiseless spectra.
}
\end{table*}

\end{appendix}

\end{document}